\documentclass[usenatbib,fleqn]{mnras}

\pdfoutput=1

\usepackage{amsmath}
\usepackage{amssymb}
\usepackage{bm}
\usepackage{color}
\usepackage{graphicx}
\usepackage{hyperref}
\usepackage[all]{hypcap}
\usepackage{longtable}
\usepackage{multicol}
\usepackage{multirow}
\usepackage{natbib}
\usepackage{flushend}
\usepackage{txfonts}
\usepackage[capitalize,nameinlink]{cleveref}

\creflabelformat{equation}{#2#1#3}
\crefrangelabelformat{equation}{#3#1#4 to #5#2#6}

\hypersetup{colorlinks,
  linkcolor=red,
  filecolor=red,
  urlcolor=magenta,
  citecolor=blue}

\def\awise{{\sc Astro-WISE}}

\def\lensfit{\emph{lens}fit}
\def\Lensfit{\emph{Lens}fit}

\def\rsat{r_{\rm sat}}
\def\Rsat{R_{\rm sat}}
\def\sex{{\sc SExtractor}}
\def\sqdeg{deg$^2$}

\def\aap{A\&A}
\def\aaps{A\&AS}
\def\aj{AJ}
\def\apj{ApJ}
\def\apjs{ApJS}
\def\araa{ARA\&A}
\def\jcap{JCAP}
\def\mnras{MNRAS}
\def\nat{Nature}
\def\pasp{PASP}

\newcommand\affil[1]{$^{#1}$}
\def\leiden{1}
\def\ucl{2}
\def\bonn{3}
\def\liverpool{4}
\def\aao{5}
\def\monash{6}
\def\edinburgh{7}
\def\uwa{8}
\def\standrews{9}
\def\inaf{10}
\def\groningen{11}
\def\oxford{12}
\def\durham{13}

\title[Satellite galaxy-galaxy lensing in KiDS$\times$GAMA]{The masses of satellites in 
GAMA galaxy groups from 100 square degrees of KiDS weak lensing data}

\author[C.~Sif\'on et al.]{
        Crist\'obal~Sif\'on\affil{\leiden}\thanks{E-mail: sifon@strw.leidenuniv.nl},
        Marcello~Cacciato\affil{\leiden},
        Henk~Hoekstra\affil{\leiden},
        Margot~Brouwer\affil{\leiden},
        \newauthor
        Edo~van~Uitert\affil{\ucl,\bonn},
        Massimo~Viola\affil{\leiden},
        Ivan~Baldry\affil{\liverpool},
        Sarah~Brough\affil{\aao},
        Michael~J.~I.~Brown\affil{\monash},
        \newauthor
        Ami~Choi\affil{\edinburgh},
        Simon~P.~Driver\affil{\uwa,\standrews},
        Thomas~Erben\affil{\bonn},
        Aniello~Grado\affil{\inaf},
        Catherine~Heymans\affil{\edinburgh},
        \newauthor
        Hendrik~Hildebrandt\affil{\bonn},
        Benjamin~Joachimi\affil{\ucl},
        Jelte~T.~A.~de~Jong\affil{\leiden},
        Konrad~Kuijken\affil{\leiden},
        \newauthor
        John~McFarland\affil{\groningen},
        Lance~Miller\affil{\oxford},
        Reiko~Nakajima\affil{\bonn},
        Nicola~Napolitano\affil{\inaf},
        \newauthor
        Peder~Norberg\affil{\durham},
        Aaron~S.~G.~Robotham\affil{\uwa},
        Peter~Schneider\affil{\bonn},
        Gijs~Verdoes~Kleijn\affil{\groningen}
\\
\affil{\leiden} Leiden Observatory, Leiden University, PO Box 9513, NL-2300 RA Leiden, Netherlands\\
\affil{\ucl} Department of Physics and Astronomy, University College London, Gower Street, London 
WC1E 6BT, UK\\
\affil{\bonn} Argelander-Institut f\"ur Astronomie, Auf dem H\"ugel 71, 53121 Bonn, Germany\\
\affil{\liverpool} Astrophysics Research Institute, Liverpool John Moores University, IC2, 
Liverpool Science Park, 146 Brownlow Hill, Liverpool, L3 5RF\\
\affil{\aao} Australian Astronomical Observatory, PO Box 915, North Ryde, NSW 1670, Australia\\
\affil{\monash} Monash Centre for Astrophysics, School of Physics and Astronomy, Monash University, 
Clayton, Victoria 3800, Australia\\
\affil{\edinburgh} Scottish Universities Physics Alliance, Institute for Astronomy, University of 
Edinburgh, Royal Observatory, Blackford Hill, Edinburgh, EH9 3HJ, UK\\
\affil{\uwa} International Centre for Radio Astronomy Research (ICRAR), The University of Western 
Australia, 35 Stirling Highway, Crawley, WA 6009, Australia\\
\affil{\standrews} Scottish Universities’ Physics Alliance (SUPA), School of Physics and Astronomy, 
University of St.\ Andrews, North Haugh, St.\ Andrews, KY16 9SS, UK\\
\affil{\inaf} INAF-Osservatorio Astronomico di Capodimonte, Via Moiariello 16 80131 Napoli Italy\\
\affil{\groningen} Kapteyn Astronomical Institute, University of Groningen\\
\affil{\oxford} Department of Physics, Oxford University, Keble Road, Oxford OX1 3RH\\
\affil{\inaf} INAF - Osservatorio Astronomico di Capodimonte, Via Moiariello 16 -80131 Napoli, 
Italy\\
\affil{\durham} ICC \& CEA, Department of Physics, Durham University, South Road, Durham, DH1 3LE, 
UK
}

\pubyear{2015}

\begin{document}
\label{firstpage}
\pagerange{\pageref{firstpage}--\pageref{lastpage}}

\maketitle

\pagebreak

\begin{abstract}
We use the first 100 \sqdeg\ of overlap between the Kilo-Degree Survey (KiDS) and the Galaxy And 
Mass Assembly (GAMA) survey to determine the galaxy halo mass of $\sim$10,000 
spectroscopically-confirmed satellite galaxies in massive ($M>10^{13}h^{-1}{\rm M}_\odot$) galaxy 
groups. Separating the sample as a function of projected distance to the group centre, we jointly 
model the satellites and their host groups with Navarro-Frenk-White (NFW) density profiles, fully 
accounting for the data covariance. The probed satellite galaxies in these groups have total masses 
$\log M_{\rm sub}/(h^{-1}{\rm M}_\odot)\approx11.7-12.2$ consistent across group-centric distance 
within the errorbars. Given their typical stellar masses, $\log M_{\rm \star,sat}/(h^{-2}{\rm 
M}_\odot)\sim10.5$, such total masses imply stellar mass fractions of $M_{\rm \star,sat}/M_{\rm 
sub}\approx0.04\,h^{-1}$. The average subhalo hosting these satellite galaxies has a mass $M_{\rm 
sub}\sim0.015M_{\rm host}$ independent of host halo mass, in broad agreement with the expectations 
of structure formation in a $\Lambda$CDM universe.
\end{abstract}


\section{Introduction}

Following a hierarchical build-up, galaxy groups grow by accretion of smaller groups and isolated 
galaxies. Tidal interactions tend to transfer mass from infalling galaxies to the (new) host 
group, with the former becoming group satellites. The favoured cosmological scenario posits that 
galaxies are embedded in larger dark matter haloes, with masses that largely exceed the stellar 
masses, a conclusion supported by a variety of observations \citep[see, e.g., the reviews 
by][]{trimble87,einasto13}. Accordingly, satellite galaxies are hosted by `subhaloes,' whose masses 
and distribution contain information on the properties of dark matter itself 
\citep[e.g.,][]{libeskind13}.

Because dark matter is (at least to a good approximation) 
dissipationless and baryons are not, it is subject to stronger tidal disruption than the baryonic 
component: energy losses cause baryons to sink to the centre of the potential more efficiently 
than dark matter, and therefore baryons are more resistant to tidal disruption \citep{white78}. 
This latter fact produces a unique prediction of the dark matter hypothesis: a satellite galaxy 
will be preferentially stripped of its dark, rather than stellar, matter. Thus, tidal stripping can 
be observed by comparing the total and stellar masses of satellite galaxies, such that galaxies 
accreted earlier have smaller mass-to-light ratios than galaxies accreted recently or (central) 
galaxies that have not been subject to tidal stripping by a larger host \citep[e.g.,][]{chang13b}.

Numerical simulations predict that tidal stripping is stronger within more centrally 
concentrated host haloes, and is more severe for more massive satellites 
\citep[e.g.,][]{tormen98,taffoni03,contini12}. Different infall timescales and concentrations 
induced by baryons (compared to dark matter-only simulations) can alter both the radial 
distribution and density profiles of subhaloes, consequently affecting tidal stripping in a 
radially-dependent manner \citep{romanodiaz10,schewtschenko11}, although this baryon-induced radial 
dependence could plausibly be (partially) compensated by feedback from active galactic nuclei 
\citep[AGN;][]{romanodiaz10}.

Observationally, the primary difficulty lies in estimating the total masses of satellite galaxies.
Weak gravitational lensing is currently the only option available to measure the total mass of 
statistical samples of galaxies. So-called (weak) galaxy-galaxy lensing provides a direct 
measure of the masses of lensing galaxies through the observation of their distortion of the images 
of background galaxies, without assumptions about the dynamical state of the system 
\citep[e.g.,][]{brainerd96,courteau14}. Weak lensing is an intrinsically statistical observational 
measurement: outside the strong lensing regime (typically a few tenths of arcsecond) the 
distortion induced in each background galaxy is much smaller than the typical galaxy ellipticity. 
Such measurements require high-quality multi-colour images that allow both accurate shape 
measurements and photometric redshift determination of faint, distant background sources. 
Measuring the lensing signal around satellite galaxies \citep[hereafter `satellite lensing', see, 
e.g.,][]{yang06} is particularly challenging because of i) the small relative contribution of the 
satellite galaxy to the lensing signal produced by the host galaxy group; ii) source blending at 
small separations, which hampers our ability to measure shapes reliably (and which is enhanced in 
high-density regions); and iii) particular sensitivity to contamination by field 
galaxies. This latter point is critical: since the dark matter haloes around satellite galaxies are 
expected to be stripped, isolated galaxies will significantly contaminate the lensing signal since 
they are not stripped, thus complicating a meaningful interpretation of the signal. Therefore, 
satellite lensing requires a clean sample of satellite galaxies to allow a proper interpretation of 
the signal. Satellite galaxies can usually be identified easily in massive galaxy clusters with 
high purity by use of, for instance, the red sequence \citep[e.g.,][]{rozo15,sifon15}, which in 
principle requires only two-band photometry. Indeed, most satellite lensing measurements so far 
have concentrated on massive galaxy clusters with deep Hubble Space Telescope (HST) observations in 
which bright cluster members can be easily identified 
\citep{natarajan02,limousin07,natarajan09}, sometimes with the aid of strong lensing 
measurements \citep{natarajan07}. Some of these studies have claimed detections of satellite 
truncation, but it seems likely that they are mostly attributable to the parameterization of 
subhalo density profiles rather than direct detections \citep{pastormira11}.

Because galaxy groups have fewer satellites than massive clusters, lensing measurements of galaxy 
group satellites require larger samples and have only been possible thanks to recent large optical 
surveys with high image quality. Furthermore, because the red sequence is generally not so well 
established in galaxy groups compared to galaxy clusters, accurate group membership determination 
requires high-completeness spectroscopic observations. Lacking such data, most measurements in 
galaxy groups to date have relied on more indirect means of estimating subhalo masses. 
\cite{gillis13} used an optimized density estimator on galaxies selected from Canada-France-Hawaii 
Telescope Lensing Survey \citep[CFHTLenS,][]{heymans12,erben13} and showed that the lensing signal 
of galaxies in high-density environments is inconsistent with the predictions of a model that does 
not include halo stripping, providing indirect evidence for tidal stripping in galaxy groups. Such 
differentiation was only possible because their high-density environment galaxies were mostly 
satellites, due to their carefully calibrated density estimator. Recently, \cite{li14} presented the 
first direct detection of the lensing signal from satellite galaxies in galaxy groups. They took 
advantage of the overlap between deep imaging from the CFHT-Stripe82 Survey \citep[CS82, 
e.g.,][]{comparat13} and the Sloan Digital Sky Survey \citep[SDSS,][]{york00} Data Release 7 
\citep{abazajian09} spectroscopic catalogue. \cite{yang07} used this SDSS catalogue to construct a 
clean galaxy group catalogue with centrals and satellites identified individually; although 
\cite{li14} had only $\sim$1,000 lens galaxies, their sample was essentially free of contamination 
by central galaxies. This allowed them to use weak lensing to directly measure the masses of 
satellites in galaxy groups for the first time, albeit with limited constraining power.

In this paper we present a direct measurement of the lensing signal from satellite galaxies in 
galaxy groups by combining a sample of spectroscopically confirmed galaxy groups from the Galaxy 
And Mass Assembly survey \citep[GAMA,][]{driver11}, and background galaxies with high-quality shape 
measurements from the Kilo-Degree Survey \citep[KiDS,][]{dejong13,kuijken15}. We use these 
measurements to constrain the masses of satellite galaxies as a function of projected distance from 
the group centre. By converting, in an average sense, these projected distances into 3-dimensional 
distances, we can study the evolution of satellite masses as they fall into galaxy groups.

This paper is organized as follows. In \Cref{s:samples} we describe the galaxy samples we use as 
lenses and lensed background sources. In \Cref{s:lensing} we summarize the measurement of 
galaxy-galaxy lensing and describe our modelling of satellites and their host groups. We present 
our results in \Cref{s:results} and summarize in \Cref{s:conclusions}. We adopt a flat $\Lambda$CDM 
cosmology with $\Omega_{\rm m}=0.315$, consistent with the latest cosmic microwave background 
measurements \citep{pcp15}, and $H_0=100h\,\mathrm{km\,s^{-1}Mpc^{-1}}$. We explicitly 
include the dependence on $h$ where appropriate. Throughout we use the symbol $\langle X \rangle$ to 
refer to the median of distribution $X$.

\section{Galaxy samples}\label{s:samples}

\begin{table*}
\begin{center}
 \caption{Median properties of satellite galaxies binned by projected distance to the group centre, 
$\Rsat$. $N_{\rm sat}$ is the total number of satellites considered while $N_{\rm sat}^{\rm KiDS}$ 
is the number of satellites that fall within the 100 \sqdeg\ of KiDS imaging used in this work. 
Errorbars are 16th and 84th percentiles.}
\label{t:sample}
\begin{tabular}{c c r c c c c c c c}
\hline\hline
 Bin & $R_{\rm sat}$ range & $N_{\rm host}$ & $N_{\rm sat}$ & $N_{\rm sat}^{\rm KiDS}$ & $\langle 
N_{\rm FoF}\rangle$ & $\langle R_{\rm sat} \rangle$ & $\langle z_{\rm sat} \rangle$ & $\log\langle 
M_{\rm \star,sat} \rangle$ & $\log\langle L_{\rm host} \rangle$ \\
 & $(h^{-1}\mathrm{Mpc})$ &  &  &  &  & $(h^{-1}\mathrm{Mpc})$ &  & $(h^{-2}{\rm M}_\odot)$ & 
$(h^{-2}{\rm L}_\odot)$ \\[0.5ex]
\hline
1 & 0.05 -- 0.20 & 1263 & 3714 & 3541 & $7_{-2}^{+5}$ & $0.12_{-0.05}^{+0.05}$ & 
$0.17_{-0.09}^{+0.09}$ & $10.45_{-0.09}^{+0.26}$ & $11.10_{-0.23}^{+0.17}$ \\[0.2ex]
2 & 0.20 -- 0.35 & 1235 & 3152 & 3042 & $7_{-2}^{+5}$ & $0.25_{-0.04}^{+0.05}$ & 
$0.19_{-0.09}^{+0.10}$ & $10.51_{-0.10}^{+0.22}$ & $11.15_{-0.26}^{+0.14}$ \\[0.2ex]
3 & 0.35 -- 1.00 &  785 & 2817 & 2773 & $8_{-3}^{+7}$ & $0.43_{-0.08}^{+0.16}$ & 
$0.21_{-0.07}^{+0.10}$ & $10.66_{-0.14}^{+0.15}$ & $11.33_{-0.25}^{+0.10}$ \\[0.2ex]
\hline
\end{tabular}
\end{center}
\end{table*}

\subsection{Lens galaxies: satellites in the GAMA galaxy group catalogue}\label{s:gama}

GAMA\footnote{\url{http://www.gama-survey.org/}} is a spectroscopic survey which measured redshifts 
for 238,000 galaxies over a total of $286\,\mathrm{deg^2}$ carried out with the AAOmega 
spectrograph on the Anglo-Australian Telescope (AAT). GAMA is 98\% spectroscopically complete down 
to $m_r=19.8$ even in the most crowded regions \citep{baldry10,driver11,liske15}. Here we use data 
over three different regions on the sky, centred at right ascensions 9h, 12h and 15h (the G09, G12 
and G15 fields), which overlap with SDSS data. Below we briefly describe the GAMA galaxy group 
sample constructed by \cite{robotham11}, who discuss the properties, possible systematics, and 
limitations of the catalogue in greater detail. Galaxy photometric properties such as luminosity 
and stellar mass are measured from the five-band optical SDSS imaging. In particular, we use the 
stellar masses derived by \cite{taylor11} by fitting \cite{bruzual03} synthetic stellar spectra to 
the broadband SDSS photometry.

The GAMA galaxy group catalogue was constructed using a 3-dimensional Friends-of-Friends (FoF) 
algorithm, linking galaxies in projected and line-of-sight separations. We use version 7 of the 
group catalogue (G$^3$Cv7), which contains 23,838 groups with $N_{\rm FoF}\geq2$, where $N_{\rm 
FoF}$ is the number of spectroscopic members grouped together by the FoF algorithm (each group has 
$N_{\rm FoF}-1$ satellites). Group properties such as velocity dispersion and total 
luminosity\footnote{The total luminosity of a group is the sum of the luminosities of its member 
galaxies, corrected for spectroscopic incompleteness at the low-mass end \citep[see][]{robotham11}.} 
were calibrated to mock galaxy catalogues processed in the same way as the real data and were 
optimized for groups with $N_{\rm FoF}\geq5$. A visual inspection of the phase space 
(distance-velocity plane) of GAMA groups confirms that groups with $N_{\rm FoF}<5$ are 
significantly contaminated by interlopers, while member selection for groups with $N_{\rm 
FoF}\geq5$ is in better agreement with the expectation of a smooth distribution of galaxies with a 
maximum velocity that decreases with radius \citep[e.g.,][]{mamon10}. We therefore restrict our 
study to groups with $N_{\rm FoF}\geq5$, in the 68.5 \sqdeg\ of unmasked area overlapping with the 
first release of KiDS lensing data (see \Cref{s:kids}). In all, we use 9683 satellites hosted by 
1467 different groups\footnote{This is the total number of satellites considered in this work, and 
includes satellites that do not fall within the currently available KiDS data but reside in a group 
which is less than $2\,h^{-1}{\rm Mpc}$ away from the center of the closest KiDS field. 9357 (97\%) 
of these satellites fall within the KiDS footprint (see \Cref{t:sample}).}. These are the same 
groups used by \cite{viola15}.

\cite{robotham11} identified the central galaxy in each group using three definitions of group 
centre: the weighted centre of light, an iterative method rejecting the galaxy farthest away from 
the center of light until one galaxy remained (the `iterative' centre), and the brightest cluster 
galaxy (hereafter BCG). All galaxies that are not centrals are classified as satellites. In most 
cases ($\sim$90\%) the iterative central galaxy coincides with the BCG, while the centre of light 
is more discrepant. \cite{viola15} performed a detailed analysis of the lensing signal of GAMA 
groups comparing the different centre definitions and confirm the results of \cite{robotham11}: the 
BCG and the iterative centre both represent the group centre of mass to a good degree, while the 
centre of light is a very poor indicator of the group centre. In this work we use the 
central-satellite classification that uses the BCG as the central, and therefore measure the 
lensing signal around all group members except the BCGs.

\subsection{Lensed background sources: the Kilo-Degree Survey}\label{s:kids}

KiDS\footnote{\url{http://kids.strw.leidenuniv.nl/}} is an ESO Public Survey being conducted with 
the 2.6 m VLT Survey Telescope (VST) in Cerro Paranal, Chile, which surveys the sky in the $ugri$ 
bands. Each 1 \sqdeg\ pointing is observed four times (`exposures') in the $u$-band and five times 
in the other bands. Upon completion, KiDS will cover 1,500 \sqdeg: half of the survey area will be 
on a 9$^\circ$-wide patch around the celestial equator and the other half on a similarly-shaped 
region around a declination of $-31^\circ$ \citep{dejong13}. In total, KiDS overlaps with four GAMA 
patches: three in the equator (the three used in this work) and one in the south (G23), for a total 
of 240 \sqdeg. In this work, we use an unmasked area of 68.5 \sqdeg\ over $\sim100$ \sqdeg\ of 
overlap currently available \citep{dejong15}.

KiDS data were reduced using two different pipelines: a reduction based on \awise\ 
\citep{mcfarland13} used to measure Gaussian-weighted aperture photometry \citep{kuijken08} and 
photometric redshifts with the Bayesian Photometric Redshift (BPZ) code \citep{benitez00}, and a 
{\sc theli} reduction \citep{erben13} used to measure galaxy shapes with \lensfit\ 
\citep{miller07,miller13,kitching08}. We briefly describe each in the following and refer to 
\cite{dejong15} and \cite{kuijken15} for details, including tests of systematic effects on shape 
measurements and photometric redshifts.

\subsubsection{Photometric redshifts}\label{s:photoz}

Photometric redshifts use  the coadded images from the KiDS public data releases DR1 and DR2 
\citep{dejong15} as input. These were processed using a pipeline largely based on the \awise\ 
optical pipeline \citep{mcfarland13} which includes crosstalk and overscan corrections, flat 
fielding, illumination correction, satellite track removal and background subtraction, plus masking 
for bad pixels, saturation spikes and stellar haloes. A common astrometric solution was calculated 
per filter using a second-order polynomial. Individual exposures were regridded and co-added using a 
weighted mean procedure. Photometric zero-points were first derived per CCD by comparing nightly 
standard star observations to SDSS DR8 \citep{aihara11} and zero-point offsets were subsequently 
applied to the $gri$ data, based on a comparison of the photometry between the CCDs in the five 
exposures. This yields a homogeneous photometry over 1 \sqdeg.

The point spread function (PSF) of the stacked images was homogenized by convolving them with a 
Gaussian kernel with varying width, such that each resulting image has a circular, Gaussian PSF 
with constant width across the field of view. A `Gaussian Aperture and PSF' 
\citep[GAaP,][]{kuijken08} photometry can be obtained such that the resulting aperture photometry 
is independent of seeing \citep[see Appendix A of][]{kuijken15}. The flux of a galaxy can then be 
measured consistently within the same physical aperture in all bands, which is necessary for 
unbiased galaxy colour estimates.

GAaP photometry was finally compared to SDSS photometry in order to obtain an absolute 
photometric calibration. Photometric redshifts were estimated using GAaP magnitudes with BPZ, 
following \cite{hildebrandt12}. \cite{kuijken15} compared the photometric redshifts to $\sim$17,000 
spectroscopic redshifts in the zCOSMOS \citep{lilly07} and ESO/GOODS \citep{vanzella08,balestra10} 
surveys. They found that the peak of the posterior distribution, $z_B$, is biased by less than 2\% 
in the range $0.005<z_B<1.0$. However, for lenses at $z_l\lesssim0.3$, as in our case, the lensing 
efficiency (cf.\ \Cref{eq:Sigma_c}) does not vary significantly for sources beyond $z_s=0.5$. In 
order to have a larger number of sources for which to measure shapes, we therefore use all galaxies 
in the range $0.005<z_B<1.2$. In the context of the CFHTLenS survey, \cite{benjamin13} have shown 
that the stacked photometric redshift posterior distribution, $p(z)$, estimated by 
\cite{hildebrandt12} in this $z_B$ range is a fair representation of the true (i.e., spectroscopic) 
redshift distribution. We therefore use the full $p(z)$ in our lensing analysis (see 
\Cref{s:lensing}).

\subsubsection{Shape measurements}\label{s:shapes}

The $r$-band data were also reduced with the {\sc theli} pipeline \citep{erben13}, independently of 
the \awise\ pipeline, in order to measure the shapes of galaxies. We used only the $r$-band data 
for shape measurements, since the $r$-band observing conditions are significantly better than in 
the other three bands \citep[see][]{dejong13}; combining different bands is not expected to result 
in a useful improvement in shape measurements. We used \sex\ \citep{bertin96} to detect objects on 
the stacked $r$-band image, and used the resulting catalogue as input to \lensfit, which is used to 
simultaneously analyze the single exposures. \Lensfit\ is a Bayesian method that returns for each 
object an ellipticity and an associated weight, $w_s$, which quantifies the measurement uncertainty 
after marginalizing over galaxy position, size, brightness, and bulge-to-disk ratio. It interpolates 
the PSF over a 2-dimensional polynomial across the image in order to estimate the PSF at the 
location of each galaxy. The number density of galaxies in the unmasked region that pass the 
photometric redshift cuts having $w_s>0$ is $n_{\rm gal}=8.88\,\mathrm{gal\,arcmin^{-2}}$ and the 
effective number of galaxies is $n_{\rm 
eff}=(\sigma_\epsilon/A)\sum_iw_{s,i}=4.48\,\mathrm{gal\,arcmin^{-2}}$ 
\citep[see][]{chang13a,kuijken15}; the root-mean-square (rms) ellipticity of galaxies is  
$\sigma_\epsilon=0.279$. We correct for \emph{noise bias}, which produces a signal-to-noise ratio 
(S/N) -dependent correction factor, $m$, between the mean ellipticity measurements and the shear 
\citep[e.g.,][see \Cref{s:lensing}]{melchior12,refregier12,viola14}, using the correction calculated 
for CFHTLenS using extensive image simulations by \cite{miller13}, which \cite{kuijken15} 
demonstrate is appropriate for the current KiDS catalogue. We also correct the galaxy shapes for an 
additive bias, $c$, introduced by imperfect PSF modelling following following \cite{heymans12}. 
See \cite{kuijken15} for details.

In performing the lensing analysis we have decided to blind ourselves to the final results. By 
doing this we ensure that the analysis does not depend on the results, and minimize the risk of 
confirmation bias. This is an especially important concern in this era of precision cosmology. At 
the start of the project we contacted an external person (unknown to all members of the KiDS 
collaboration except for the contact person), who generated three additional catalogues by 
rescaling the galaxy ellipticities by factors unknown to us. We carried out the full analysis four 
times, one for each ellipticity catalogue. Only when the team was convinced about the analysis 
\emph{carried out with the four ellipticity catalogues}, the analysis was frozen \emph{with no 
further changes to the results} and we contacted the external person again to reveal the true 
catalogue. A detailed description of the shape analysis and catalogue blinding of KiDS data is 
given in \cite{kuijken15}.

\section{Galaxy-galaxy lensing of satellite galaxies}\label{s:lensing}

Gravitational lensing produces a differential deflection of light coming from background galaxies 
when it passes through an inhomogeneous mass distribution, and most strongly along a mass 
concentration. The observable effect is a coherent distortion on both the shape and the size of 
background sources around the lens. The shape distortion, $\gamma_{\rm t}$, is referred to as 
\emph{shear}, and in the weak lensing limit is much smaller than the typical ellipticities of 
galaxies and can only be measured statistically by averaging over many background sources. The 
average tangential shear relates to the excess surface mass density (ESD) at a projected 
distance\footnote{As a convention, we list 3-dimensional distances in groups with lower case $r$, 
and distances projected in the plane of the sky with capital $R$.} $R$ of the lens, 
$\Delta\Sigma(R)$, through
\begin{equation}\label{eq:gammat}
 \Delta\Sigma(R) \equiv \bar\Sigma(<R)-\bar\Sigma(R) = \Sigma_{\rm c}\gamma_{\rm t}(R)\,,
\end{equation}
where $\bar\Sigma(<R)$ is the average surface density within $R$, $\bar\Sigma(R)$ is the average 
surface density \emph{at} $R$ (more precisely, within a thin shell $R+\delta R$) and the critical 
density, $\Sigma_{\rm c}$, is as a geometrical factor that accounts for the lensing efficiency,
\begin{equation}\label{eq:Sigma_c}
 \Sigma_{\rm c} = \frac{c^2}{4\pi G} \frac{D(z_{\rm s})}{D(z_{\rm l})D(z_{\rm l},z_{\rm s})}.
\end{equation}
Here, $D(z_{\rm l})$, $D(z_{\rm s})$, and $D(z_{\rm l},z_{\rm s})$ are the angular diameter 
distances to the lens, to the source and between the lens and the source, respectively. Therefore 
the redshifts of the lenses and sources are essential to relate the tangential distortions of the 
sources to the projected mass density of the lens.

We calculate $D(z_{\rm l})$ for each lens galaxy using its spectroscopic redshift from GAMA and 
marginalize over the full probability distribution of the photometric redshift of each background 
source, $p(z_s)$. Specifically, for every lens-source pair we calculate
\begin{equation}
 \tilde\Sigma_{{\rm c},ls}^{-1} = \frac{4\pi G}{c^2}D(z_{\rm l}) \int_{z_l}^\infty{\rm 
d}z_sp(z_s)\frac{D(z_{\rm l},z_{\rm s})}{D(z_{\rm s})}.
\end{equation}
Each lens-source pair is then assigned a weight that combines the \lensfit\ weight and the lensing 
efficiency,
\begin{equation}
 w_{ls} = w_s\tilde\Sigma_{{\rm c},ls}^{-2}.
\end{equation}
The ESD in a bin centred on a projected distance $R$ is then calculated as
\begin{equation}
 \Delta\Sigma(R) = \left(\frac{\sum_{ls}w_{ls}\epsilon_{\rm t}\tilde\Sigma_{{\rm c},ls}} 
{\sum_{ls}w_{ls}}\right)\frac1{1+K(R)}
\end{equation}
where the sum is over all lens-source pairs in the radial bin, $\epsilon_{\rm t}$ is the tangential 
component of the ellipticity of each source around each lens, and
\begin{equation}
 K(R) = \frac{\sum_{ls}w_{ls}m_s}{\sum_{ls}w_{ls}} \simeq0.1,
\end{equation}
where $m$ is the multiplicative correction for noise bias \citep{miller13,kuijken15}.

The ESD around a satellite galaxy at a projected distance $\Rsat$ from the group centre, 
$\Delta\Sigma_{\rm sat}(R \vert \Rsat)$, is given by
\begin{equation}\label{eq:DeltaSigma}
 \Delta\Sigma_{\rm sat}(R \vert \Rsat) = \Delta\Sigma_{\rm sub}(R) + \Delta\Sigma_{\rm 
host}(R \vert \Rsat)\,,
\end{equation}
where $\Delta\Sigma_{\rm sub}$ is the ESD of the subhalo in which the satellite galaxy resides and 
$\Delta\Sigma_{\rm host}$ is the ESD of the host galaxy group, measured around the satellite 
galaxy. We describe the measured satellite lensing signal in \Cref{s:signal} before discussing our 
modelling of both terms of \Cref{eq:DeltaSigma} in \Cref{s:sigma_host,s:sigma_sat}. In doing this, 
we follow the discussion by \cite{yang06}.

\subsection{The satellite lensing signal}\label{s:signal}

\begin{figure}
 \centerline{\includegraphics[width=3.4in]{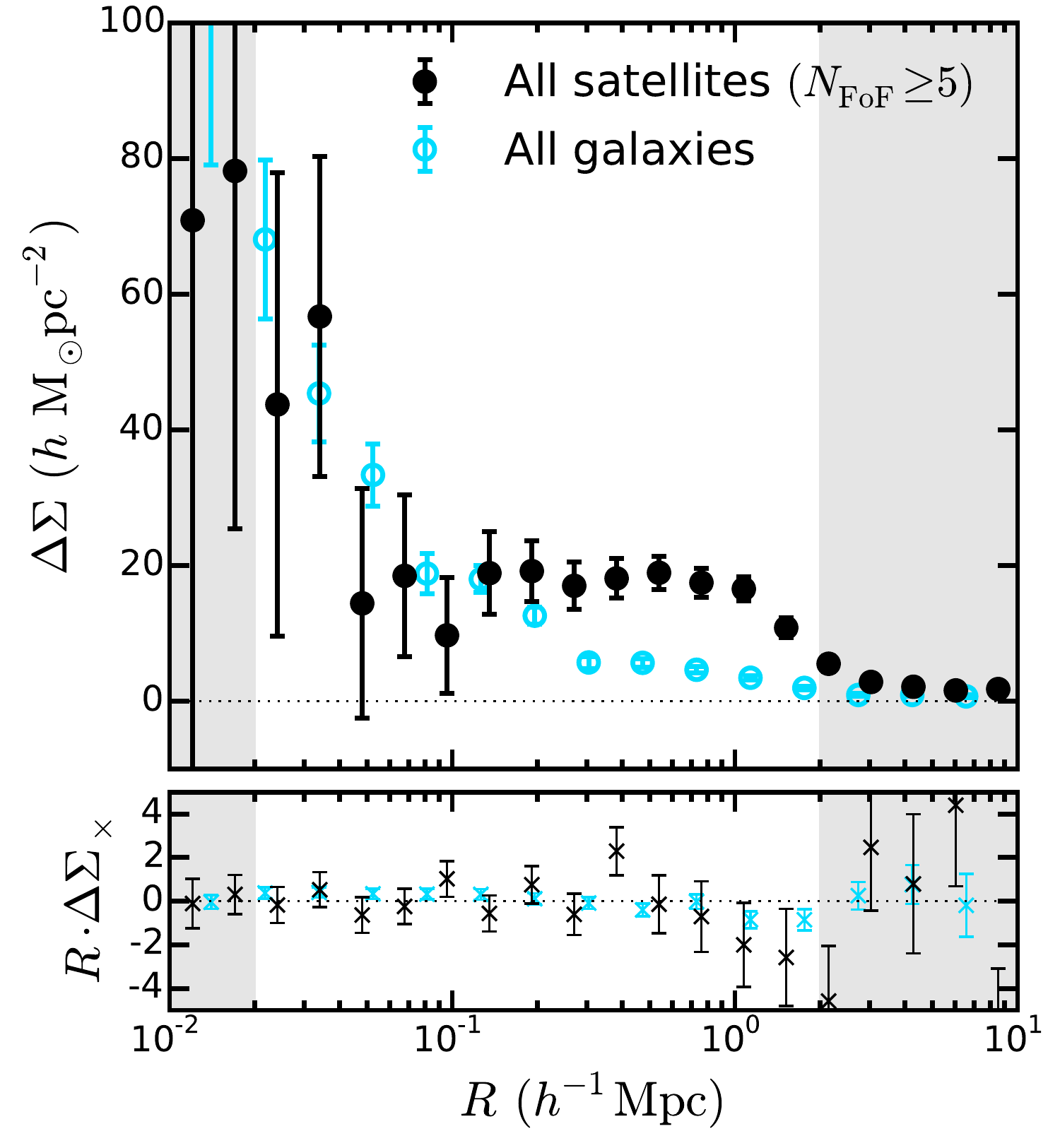}}
\caption{
\textit{Top panel:} Excess surface density around all satellites residing in groups with $N_{\rm 
FoF}\geq5$ (black points) and around all galaxies in the GAMA catalogue (cyan circles). 
\textit{Bottom panel:} corresponding cross signals, multiplied by projected separation, $R$, to 
make the errorbars of comparable size throughout the radial range (units are omitted for clarity). 
Dotted horizontal lines in both panels show $\Delta\Sigma=0$. We used different bins to measure the 
signal of each sample for clarity. The grey bands show projected separations that are not used in 
our analysis.}
\label{f:esd_all}
\end{figure}

\begin{figure*}
 \centerline{\includegraphics[width=2.2in]{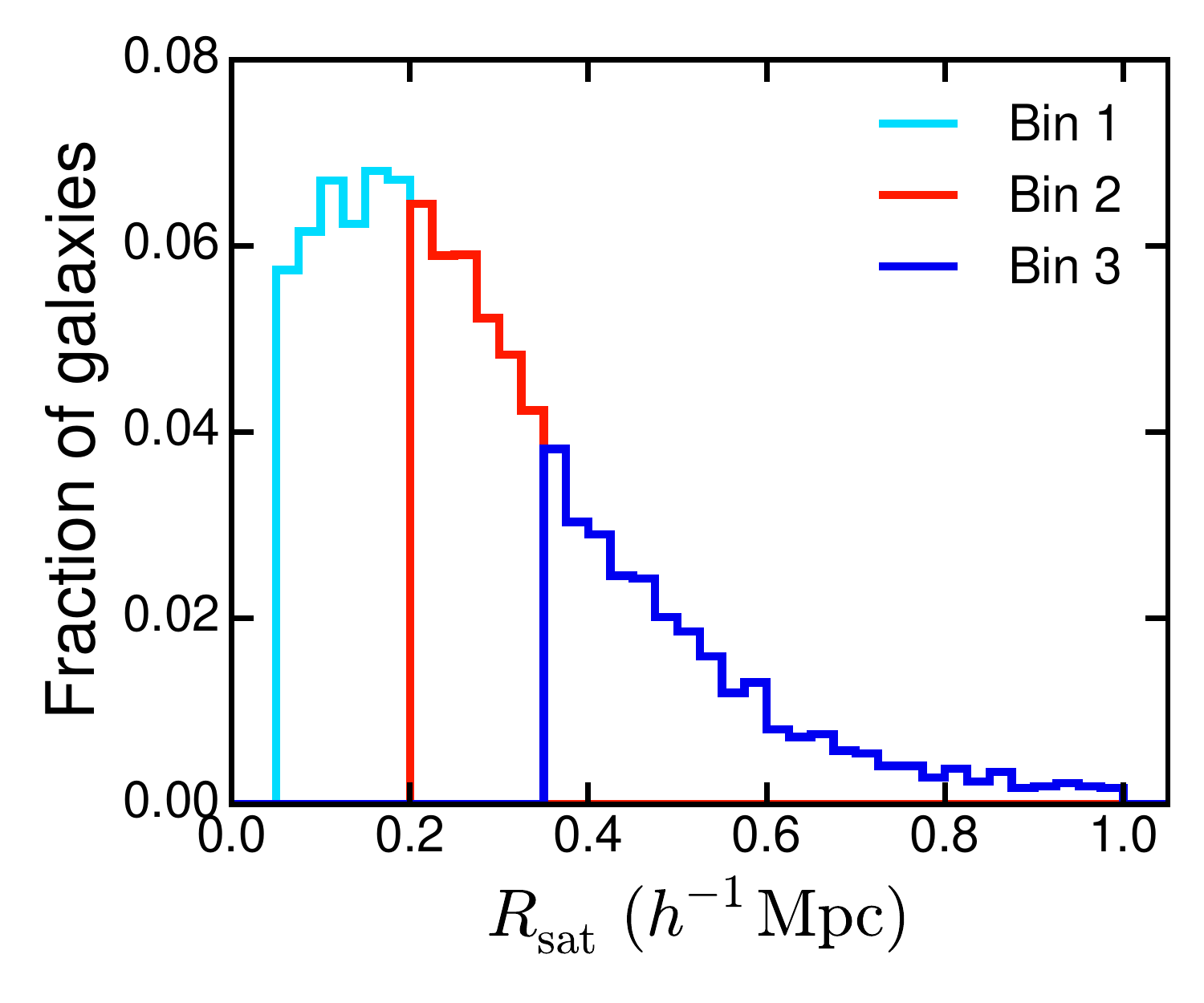}
             \includegraphics[width=2.2in]{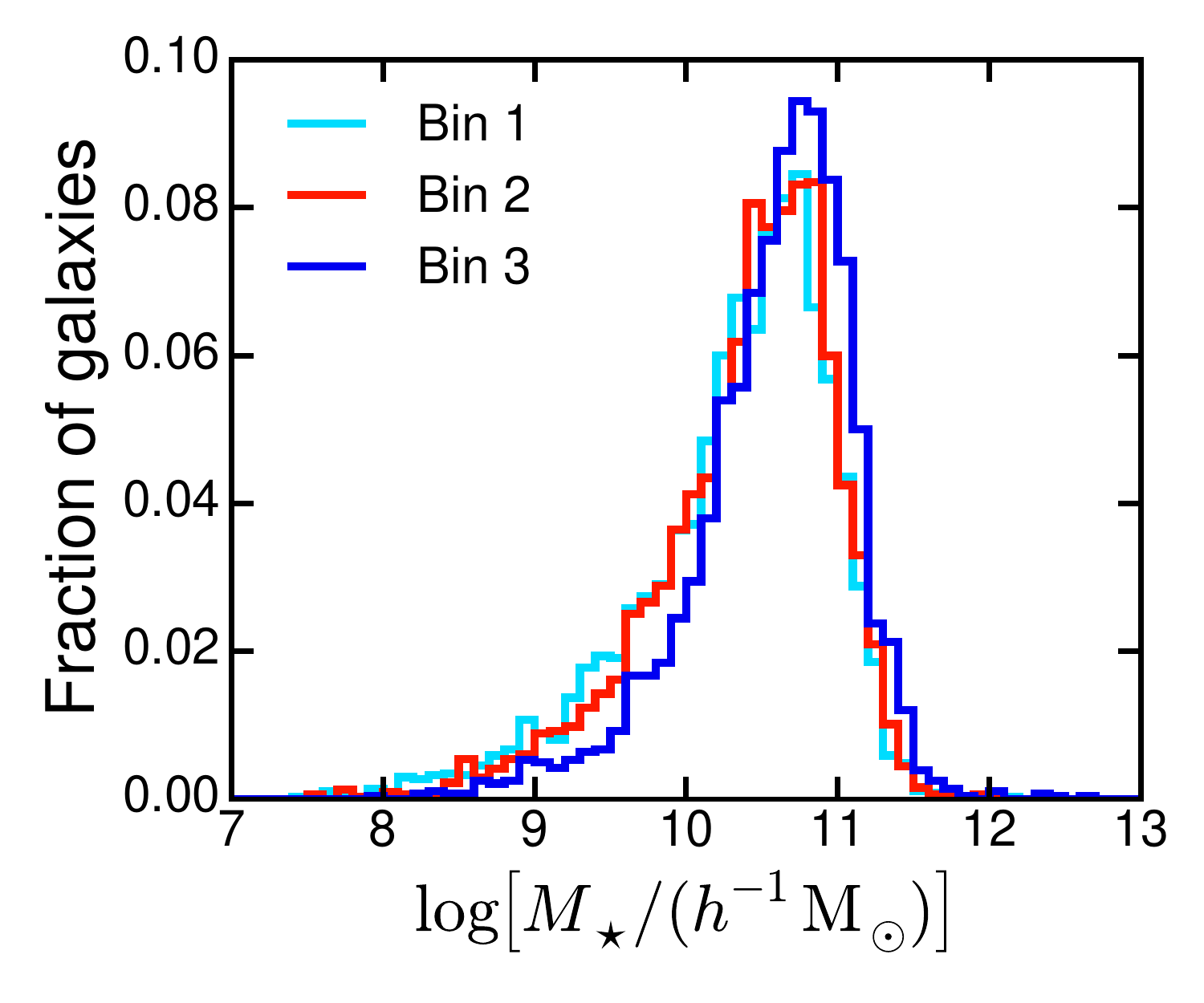}}
 \centerline{\includegraphics[width=2.2in]{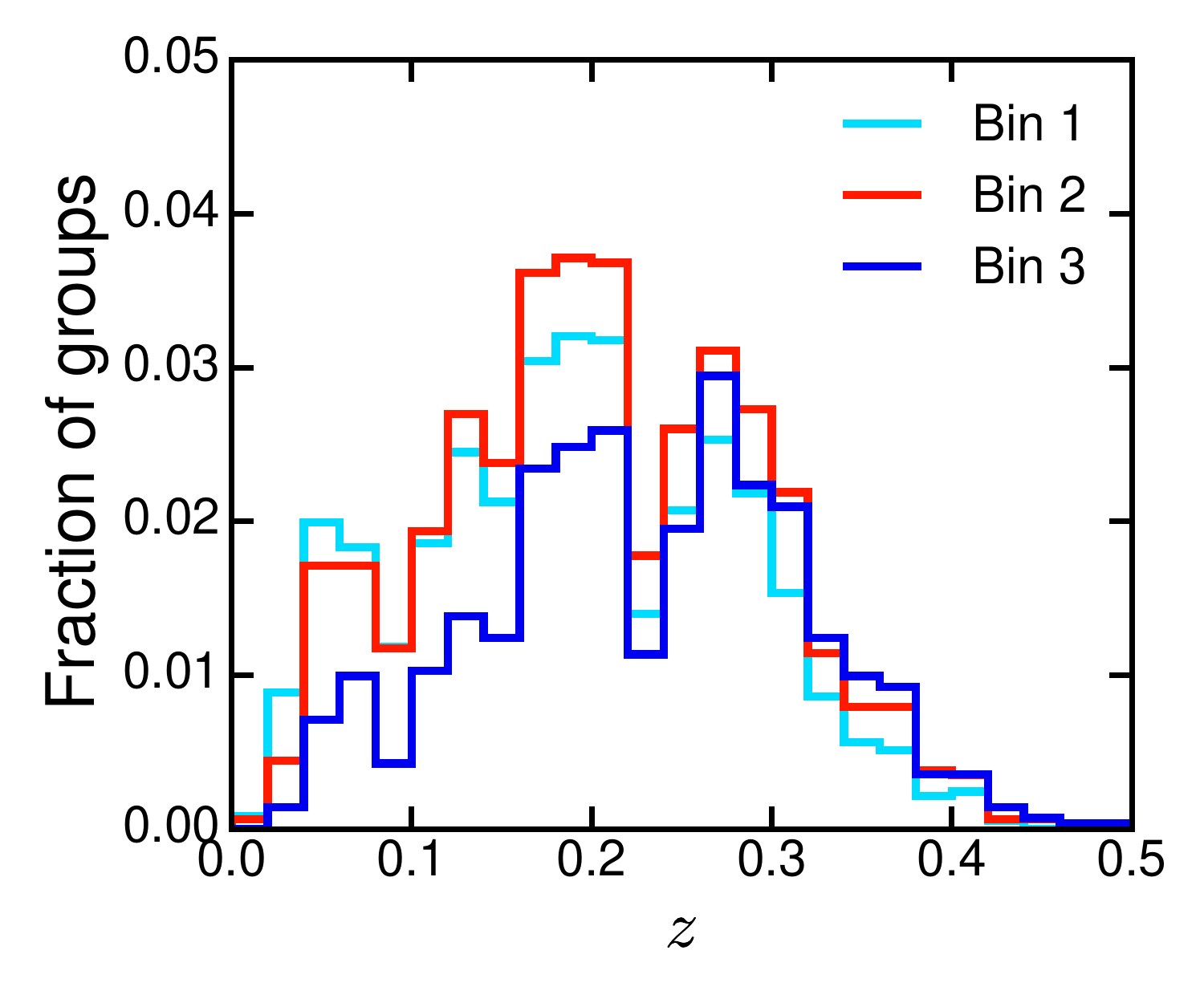}
             \includegraphics[width=2.2in]{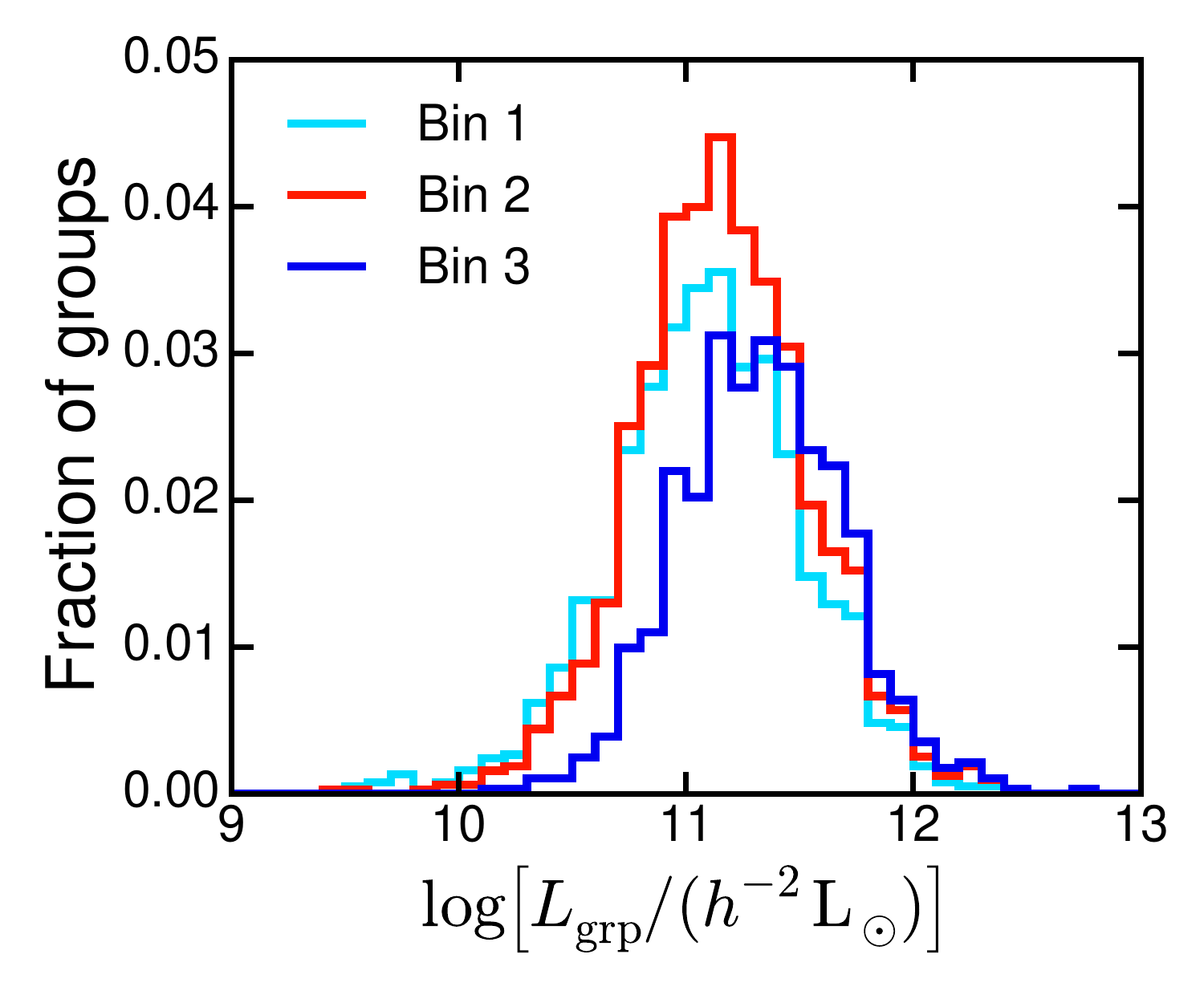}}
\caption{\textit{Top:} Satellite distributions of distance to the BCG (\textit{left}) and stellar 
mass (\textit{right}); \textit{bottom:} Group distributions of redshift (\textit{left}) and total 
luminosity (\textit{right}); for the radial bins defined in \Cref{t:sample}. Note that each group 
can contribute to more than one bin in the lower panels.}
\label{f:histograms}
\end{figure*}

We show in \Cref{f:esd_all} the stacked ESD of all 9683 satellites residing in groups with $N_{\rm 
FoF}\geq5$. We also show the ESD around all galaxies in the GAMA catalogue, which is dominated by 
(central) field galaxies \citep{robotham11}. The lensing signal around the two samples is 
qualitatively different. In terms of \Cref{eq:DeltaSigma}, the ESD of central galaxies can be 
described by $\Delta\Sigma_{\rm host}(R \vert \Rsat=0)$ alone (see \cite{vanuitert15} for a 
detailed comparison of the lensing signal of different lens samples). The bottom panel of 
\Cref{f:esd_all} shows $\Delta\Sigma_\times$, which is defined analogously to \Cref{eq:gammat} using 
the shear measured at 45$^\circ$ rotations from the direction tangential to the lens. 
$\Delta\Sigma_\times$ should be consistent with zero because of parity symmetry \citep{schneider03}, 
and therefore serves as a check for systematic effects. As shown in \Cref{f:esd_all}, 
$\Delta\Sigma_\times$ is consistent with zero for both samples at all lens-source separations.

Although in \Cref{f:esd_all} we show the lensing signal for separations $0.01 \leq R/(h^{-1}{\rm 
Mpc}) \leq 10$, we only use measurements of $\Delta\Sigma$ in the range $0.02 \leq 
R/(h^{-1}{\rm Mpc}) \leq 2$ in our analysis. Separations outside this range are marked in 
\Cref{f:esd_all} by grey bands. At smaller separations, blending with and obscuration by group 
members become significant and therefore the S/N is very low; at larger separations the coverage is 
highly incomplete due to the patchiness of the current KiDS data, making measurements less 
reliable. We assess the effect of the patchy coverage by measuring the lensing signal around random 
locations on the images, which should be consistent with zero. The signal is indeed consistent with 
zero for separations $R\lesssim5\,h^{-1}{\rm Mpc}$, but at separations $R\gtrsim5\,h^{-1}{\rm Mpc}$ 
the lensing signal around random points deviates significantly from zero \citep[see][]{viola15}. 
This indicates that systematic effects are affecting the shear estimation at such distances. We do 
not try to correct for such effects and instead conservatively discard measurements at separations 
$R>2\,h^{-1}{\rm Mpc}$.

The errorbars in \Cref{f:esd_all} correspond to the square root of the diagonal elements of the 
covariance matrix, described in \Cref{ap:cov}. In principle, the lensing covariance matrix includes 
contributions from shape noise and sample (`cosmic') variance. Shape noise arises because galaxies 
are intrinsically elliptic and because noise in the images introduces additional uncertainties in 
the shape measurements \citep[see, e.g.,][]{hoekstra00}, while sample variance accounts for the 
finite fraction of the sky observed. As we show in \Cref{ap:cov}, the contribution from sample 
variance can be safely neglected for our purposes and we therefore include only the contribution 
from shape noise, which can be calculated directly from the data \citep[see Section 3.4 
of][]{viola15}. In addition to the covariance between data points (as in the case of 
\Cref{f:esd_all}), we also compute the covariance between data points around lenses in different 
bins of projected distance from the group centre, $\Rsat$ (see \Cref{f:cov}).

\begin{figure*}
 \centerline{\includegraphics[width=1.7in]{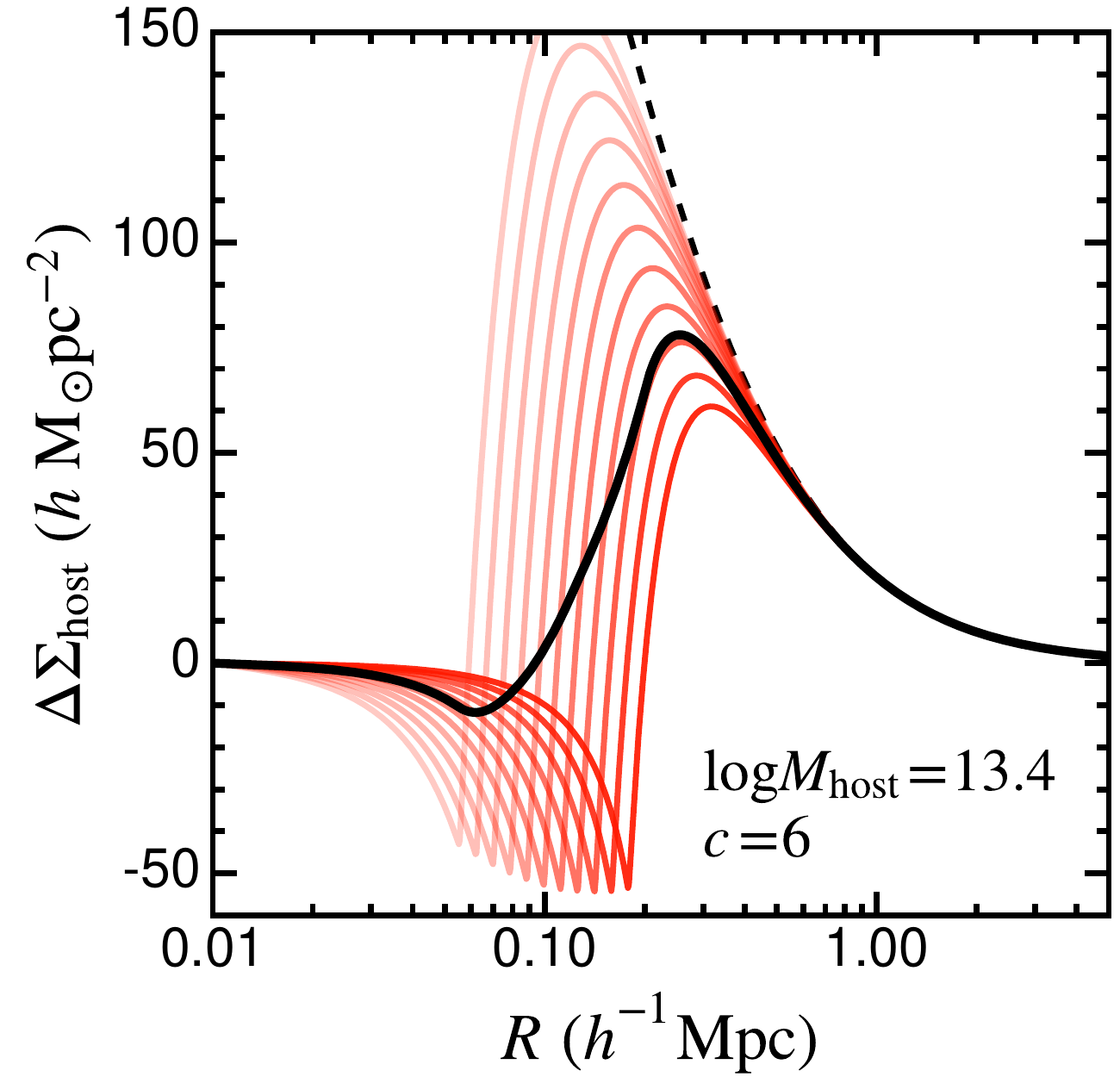}
             \includegraphics[width=1.7in]{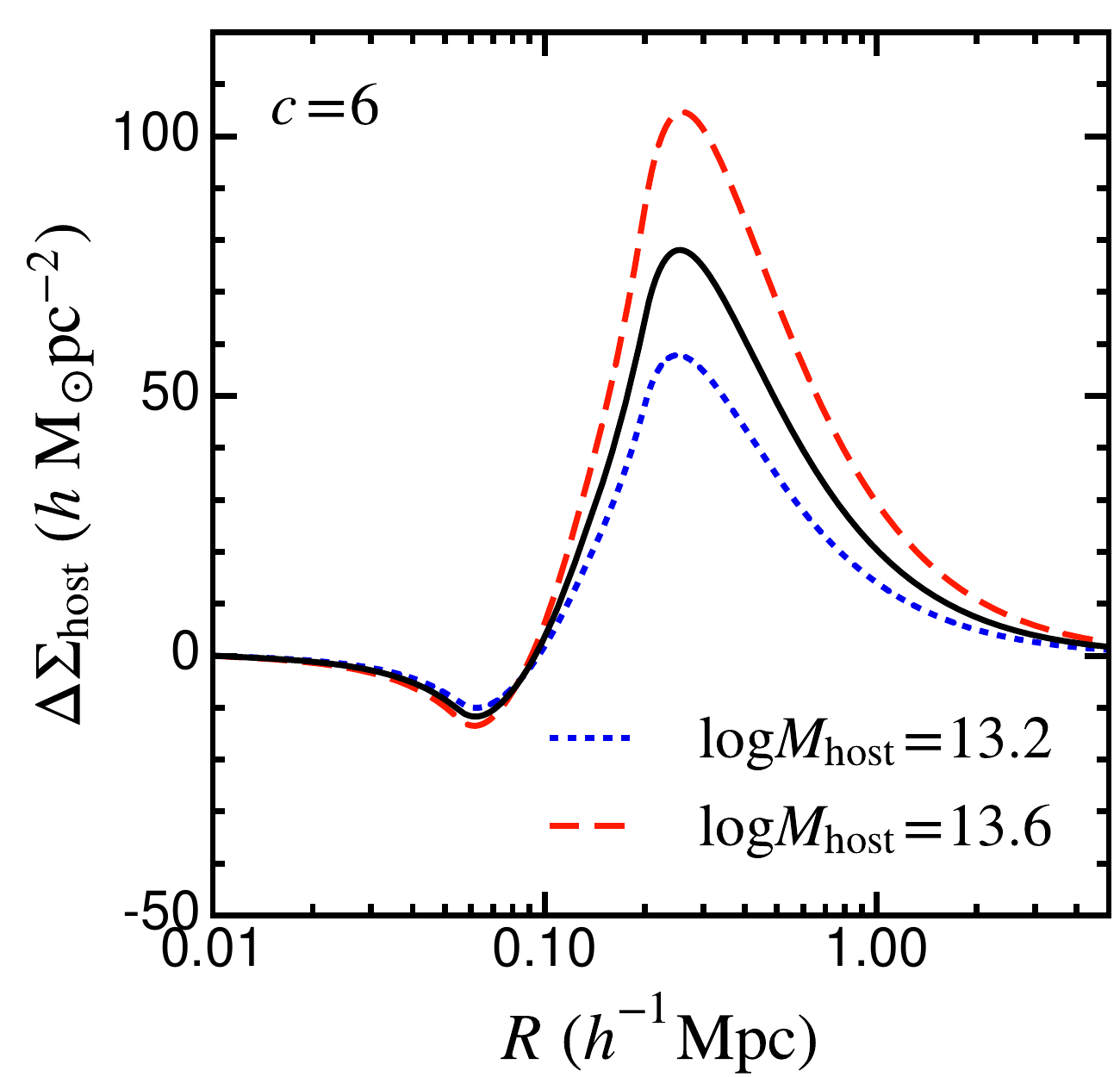}
             \includegraphics[width=1.7in]{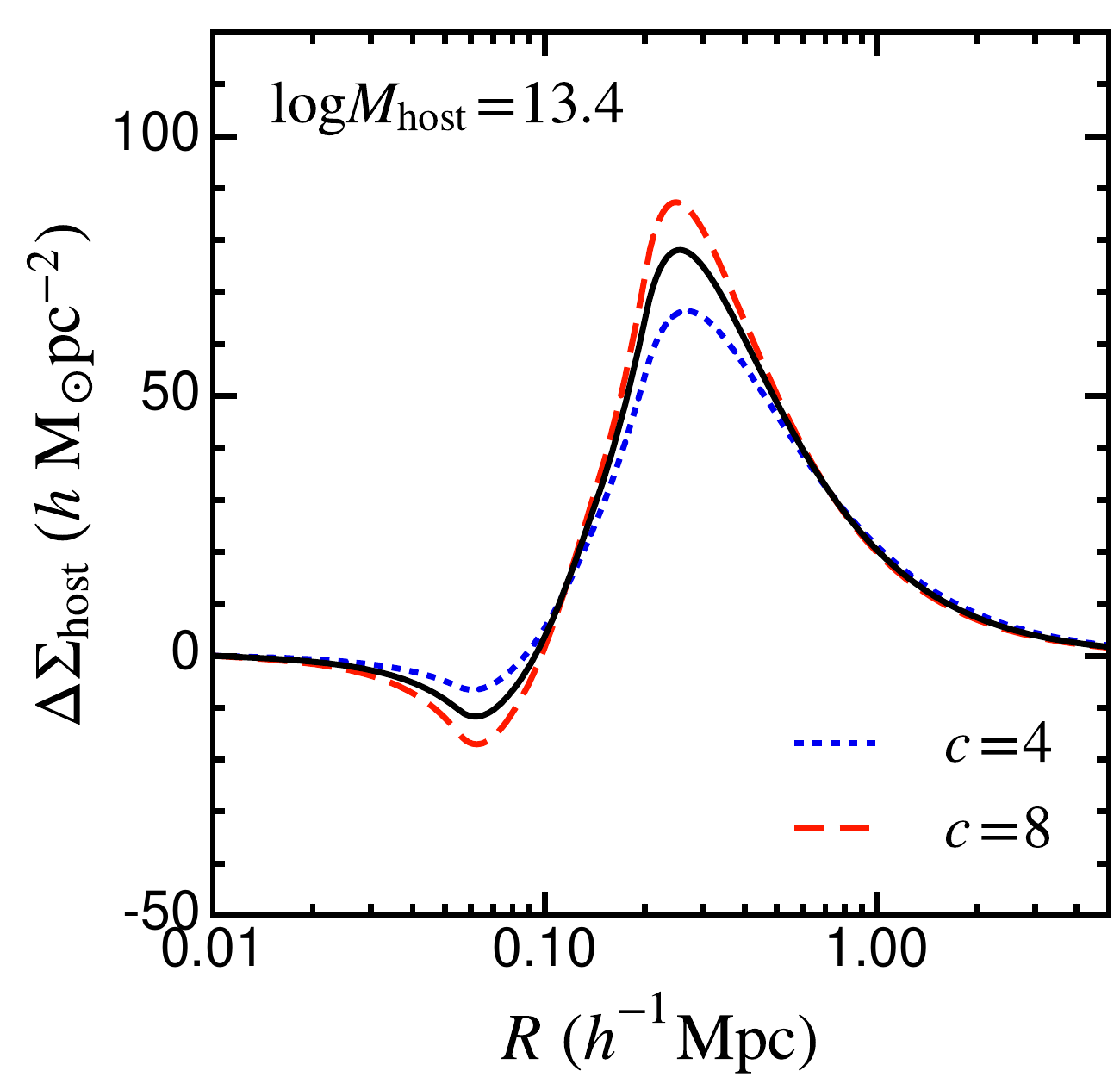}}
\caption{Illustration of the contribution from the host group, $\Delta\Sigma_{\rm host}(R\vert 
n(R_{\rm sat}))$, to \Cref{eq:DeltaSigma}.
{\it Left:} Red lines show the contribution to the signal around satellites at different 
distances from the group centre in logarithmic bins, with opacity scaling with the number density 
of objects in each bin, $n(\Rsat)$, corresponding to the cyan histogram in the top-left panel of 
\Cref{f:histograms}. The thick black line is the weighted average of the red lines 
(cf.\ \Cref{eq:esd_sum}) and represents the group contribution to the lensing signal around our 
sample of satellites with $0.05\leq\Rsat/(h^{-1}{\rm Mpc})\leq0.20$ in a group with $\log 
M_{200}=13.4$ and $c=6$, and is reproduced in the middle and right panels. The black dashed line 
shows the excess surface density of the same group when measured around the group centre.
{\it Middle:} Varying group mass at fixed concentration.
{\it Right:} Varying group concentration at fixed mass.
Note that the vertical scale in the middle and right panels is zoomed in with respect to the left 
panel. All masses are in units of $h^{-1}{\rm M}_\odot$.
}
\label{f:illustration_group}
\end{figure*}

The signal shown in \Cref{f:esd_all} has a high S/N, but its interpretation is complicated by the 
mixing of satellites with a wide range of properties. \cite{vanuitert15} use this satellite sample 
to study the stellar-to-halo mass relation by binning the sample in stellar mass and redshift. 
Here, we bin the sample by projected distance to the group centre; this binning is shown in the 
top-left panel of \Cref{f:histograms} (see also \Cref{t:sample}). We find that this particular 
binning allows us to study each bin with high enough S/N. We take the distance from the group 
centre as a proxy for time since infall to the group \citep[e.g.,][]{gao04,chang13b} and study the 
evolution of the mass in satellites as these galaxies interact with their host groups. As shown in 
\Cref{f:histograms} (top right), the three radial bins have similar stellar mass distributions, 
their medians differing by only 0.2 dex (\Cref{t:sample}). In contrast, the group redshift and 
luminosity distributions of bin 3 are different from the other two bins. Because we separate groups 
by satellite distance, we essentially split groups by size. Only the most massive (i.e., the most 
luminous) groups in the sample have satellites at $R_{\rm sat}>0.35\,h^{-1}{\rm Mpc}$. Additionally, 
because GAMA is a magnitude-limited survey, groups at high redshift are on average more luminous 
(i.e., more massive), which causes the different redshift distributions.

\subsection{Host group contribution}\label{s:sigma_host}

The average density profile of galaxy groups is well described by a Navarro-Frenk-White 
\citep[NFW,][]{navarro95} profile,
\begin{equation}\label{eq:NFW}
 \rho_{\rm NFW}(r) = \frac{\delta_c \rho_m}{(r/r_s)(1+r/r_s)^2},
\end{equation}
where $\rho_m(z)=3H_0^2(1+z)^3\Omega_M/(8\pi G)$ is the mean density of the Universe at redshift 
$z$ and
\begin{equation}\label{eq:delta_c}
 \delta_c = \frac{200}{3}\frac{c^3}{\ln(1+c) - c/(1+c)}.
\end{equation}
The two free parameters, $r_s$ and $c \equiv r_{200}/r_s$, are the scale radius and concentration 
of the profile, respectively. However, we use the concentration and the mass\footnote{Here 
$M_{200}$ is the mass within a radius $r_{200}$, which encloses a density 
$\rho(<r_{200})=200\rho_m(z)$.}, $M_{200}$, as the free parameters for convenience. We further 
assume the mass-concentration relation, $c(M,z)$, of \cite{duffy08}, allowing for a free 
normalization, $f_c^{\rm host}$. That is,
\begin{equation}\label{eq:cM}
 c(M_{200},z) = f_c^{\rm host}\left[10.14\left(\frac{M_{200}}{2\times10^{12}h^{-1}{\rm 
M}_\odot}\right)^{-0.089}\left(1+z\right)^{-1.01}\right].
\end{equation}

\begin{figure*}
 \centerline{\includegraphics[width=1.7in]{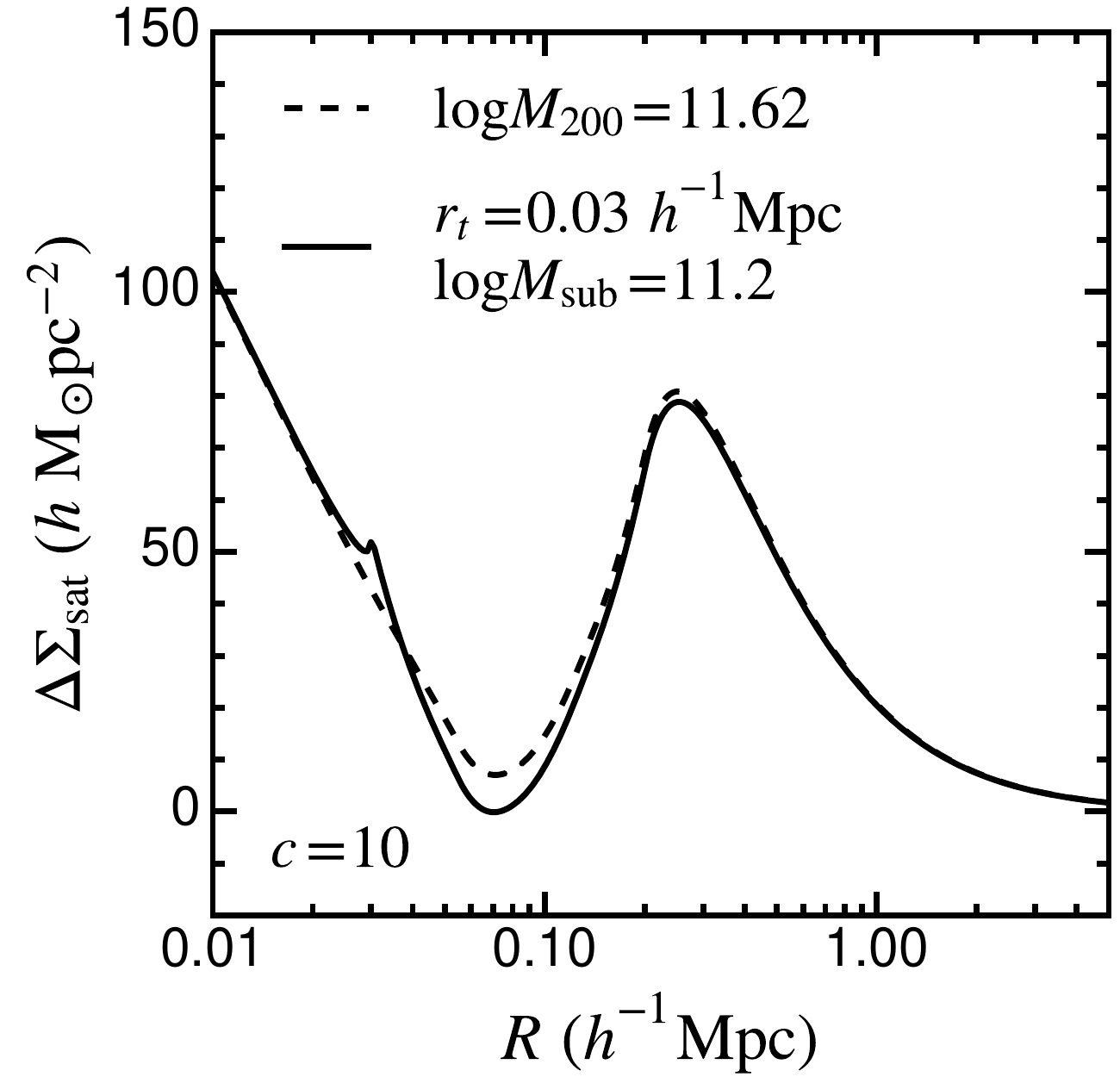}
             \includegraphics[width=1.7in]{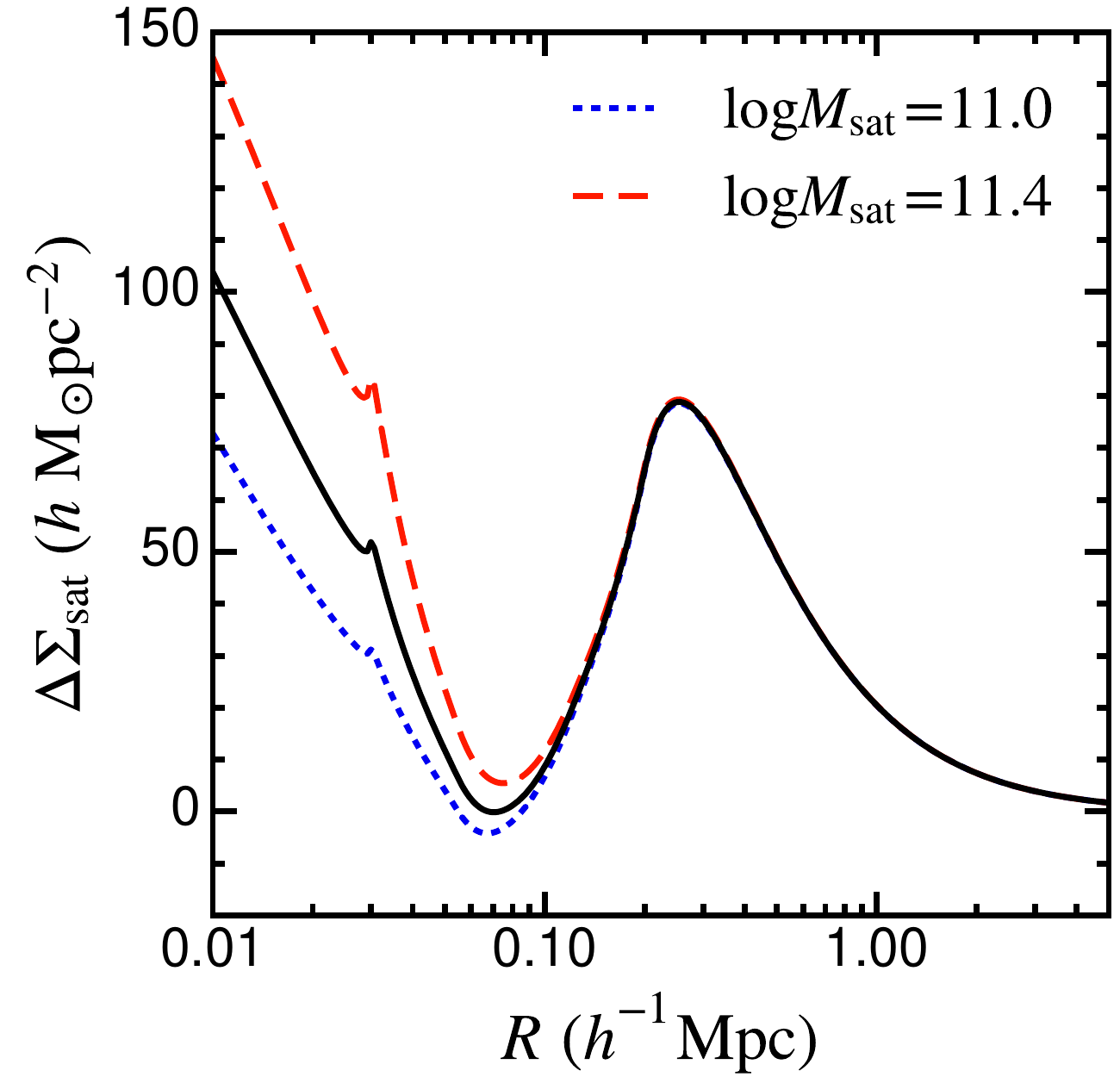}
             \includegraphics[width=1.7in]{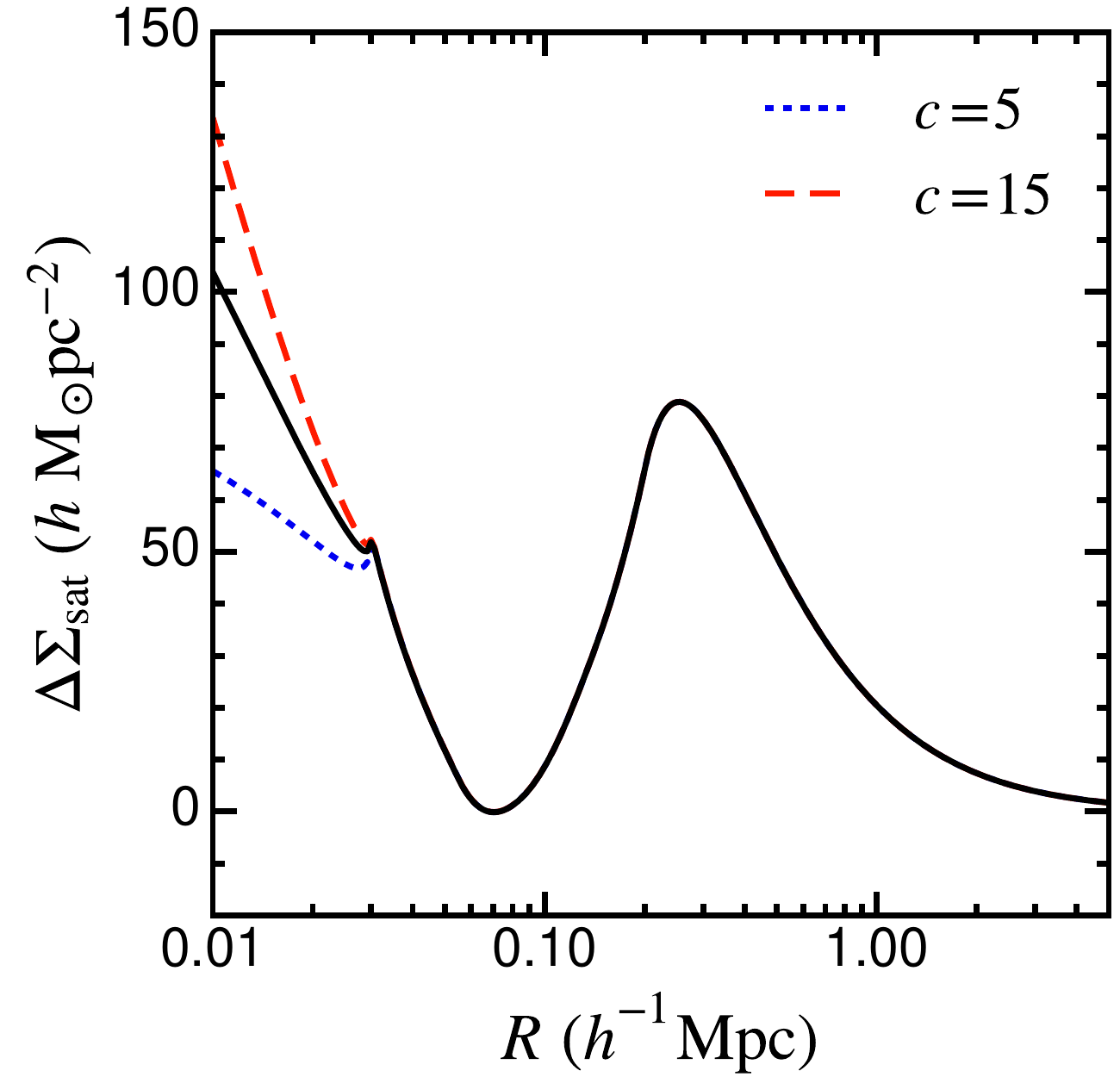}
             \includegraphics[width=1.7in]{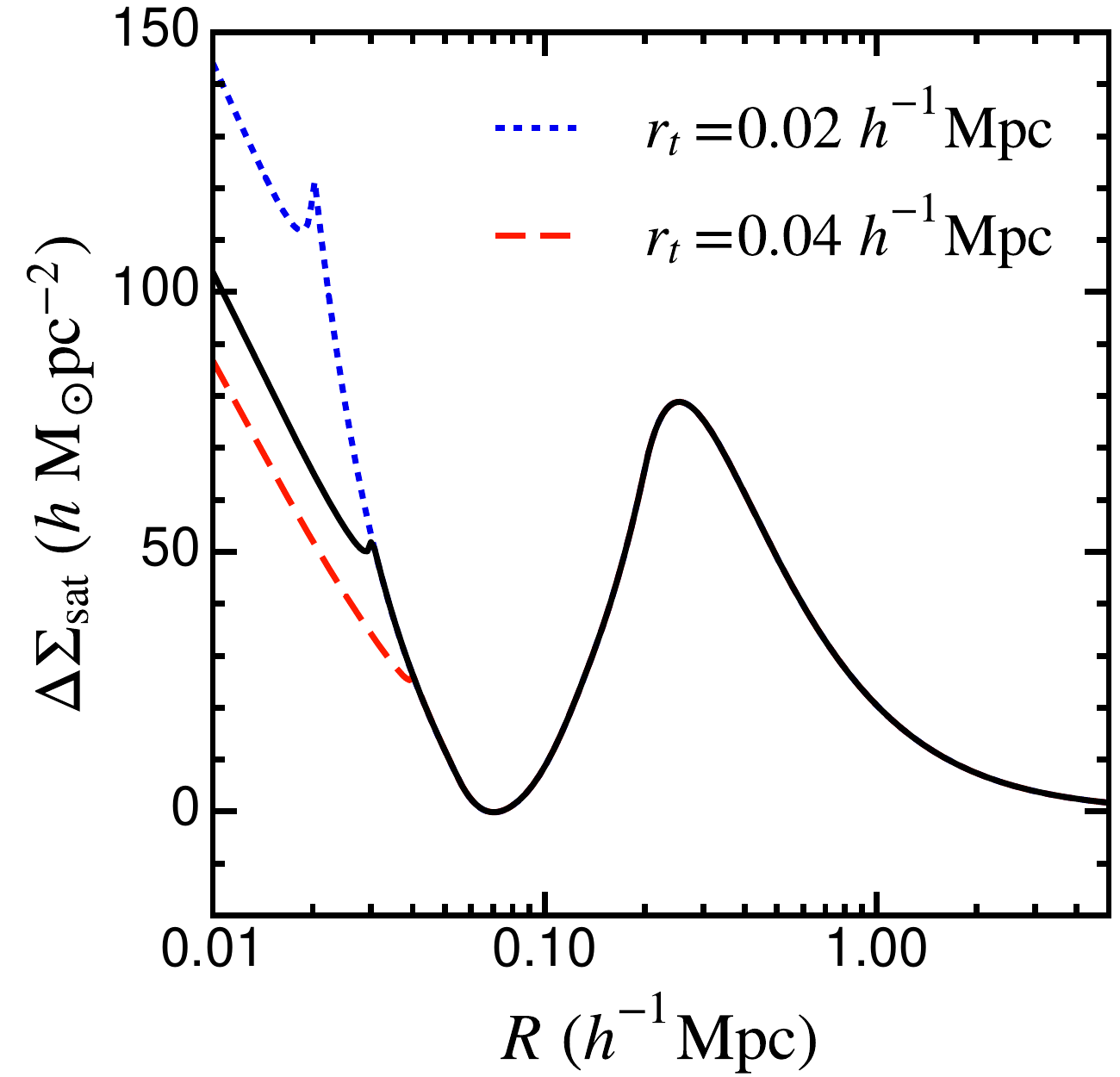}}
\caption{The satellite lensing signal, $\Delta\Sigma_{\rm sat}$, for different satellite 
properties. The group contribution is kept fixed at the fiducial value (i.e., the thick solid line) 
from \Cref{f:illustration_group}.
{\it Leftmost:} The dashed line shows the excess surface density of a NFW profile with $c=10$ and 
$\log M_{200}=11.62$. Truncating this profile through \Cref{eq:tNFW} at $r_{\rm t}=0.03\,h^{-1}{\rm 
Mpc}\approx2.6r_s$ produces the solid line, with a total mass $\log M_{\rm sub}=11.2$, which is 
reproduced in all other panels. The glitch in the solid line is produced by the sharp truncation of 
the density profile and is continuous but non-differentiable.
{\it Left-centre:} Varying $M_{\rm sub}$, keeping both $c=10$ and $r_{\rm t}=0.03\,h^{-1}{\rm Mpc}$ 
fixed.
{\it Right-centre:} Varying the concentration, keeping both $\log M_{\rm sub}=11.2$ and 
$r_{\rm t}=0.03\,h^{-1}{\rm Mpc}$ fixed.
{\it Rightmost:} Varying the truncation radius, keeping both $c=10$ and $\log M_{\rm sub}=11.2$ 
fixed. Note that the normalization of the inner profile changes because we fix the mass 
\emph{within the truncation radius}, which is itself changing.
All masses are in units of $h^{-1}{\rm M}_\odot$.
}
\label{f:illustration_satellite}
\end{figure*}

The average surface density of the host group measured at a projected distance $\Rsat$ from the 
group centre is simply the azimuthal average of $\Sigma_{\rm host}$ around the satellite,
\begin{equation}\label{eq:NFWdisplaced}
 \bar\Sigma_{\rm host}(R \vert \Rsat) = \frac1{2\pi}\int_0^{2\pi}{\rm d}\theta\,\Sigma_{\rm 
NFW}\left(\sqrt{\Rsat^2 + R^2 + 2R\Rsat\cos\theta}\right),
\end{equation}
and the contribution to the satellite ESD follows from \Cref{eq:gammat}. We use the analytical 
expression for the projected surface density of an NFW profile, $\Sigma_{\rm NFW}(R)$, derived 
by \cite{wright00}.

In reality we observe a sample of satellites at different distances to their respective group 
centres; therefore the total group contribution is
\begin{equation}\label{eq:esd_sum}
 \Delta\Sigma_{\rm host}(R\vert n(R_{\rm sat})) =
     \frac{\int_{\Rsat^{\rm min}}^{\Rsat^{\rm max}}{\rm d}
           \Rsat n(R_{\rm sat})\Delta\Sigma_{\rm host}(R \vert \Rsat)}
          {\int_{\Rsat^{\rm min}}^{\Rsat^{\rm max}}{\rm d}
           \Rsat n(R_{\rm sat})},
\end{equation}
where $n(\Rsat)$ is the number density of satellites at $\Rsat$. We use \Cref{eq:esd_sum} to model 
the host group contribution to \Cref{eq:DeltaSigma} throughout. Our implementation differs 
from that introduced by \cite{yang06} and applied by \cite{li14} in that they fit for $R_{\rm sat}$, 
whereas we use the measured separations to fix $n(R_{\rm sat})$.

We illustrate the difference between $\Delta\Sigma_{\rm host}(R\vert\Rsat)$ and $\Delta\Sigma_{\rm 
host}(R\vert n(R_{\rm sat}))$ in the left panel of \Cref{f:illustration_group}, for the innermost 
bin considered in this work (see \Cref{t:sample} and the top left panel of  \Cref{f:histograms}). 
The left panel of \Cref{f:illustration_group} shows that $\Delta\Sigma_{\rm host}(R\vert\Rsat)$ 
of a single group-satellite pair has a sharp minimum at $R=\Rsat$ where $\Sigma(R)$ is maximal and 
therefore $\Delta\Sigma<0$; $\Delta\Sigma_{\rm host}(R\vert\Rsat)$ increases abruptly further out 
and then drops back to the group's outer profile, matching the group profile measured around the 
group centre. Accounting for the distribution of group-centric distances shifts this minimum to 
$R<\langle\Rsat\rangle$, and makes both the peak and the dip significantly less pronounced; 
including the distribution of projected distances is critical to properly model the statistical 
properties of the lensing signal which cannot be captured by fitting for an average value. 
Similarly, the middle and right panels of \Cref{f:illustration_group} show the effect of different 
host masses and concentrations on $\Delta\Sigma_{\rm host}(R\vert n(R_{\rm sat}))$. A higher mass 
increases its amplitude at all scales where the host contribution dominates, whereas a higher 
concentration mostly enhances the ESD signal around the peak and produces a more pronounced dip.

Our model ignores the contribution from baryons in the central group galaxy, which are noticeable 
at scales $R<0.05\,h^{-1}{\rm Mpc}$ \citep{viola15}. Because baryons are more concentrated than 
dark matter, they can make the total density profile steeper than a pure NFW. \cite{viola15} have 
shown, however, that the amplitude of the baryonic contribution (modelled as a point mass) is not 
degenerate with any other group parameter in their halo model. Therefore we expect baryons in the 
BCG to have no impact on our results.

\subsection{Satellite contribution}\label{s:sigma_sat}

\begin{figure*}
 \centerline{\includegraphics[width=7in]{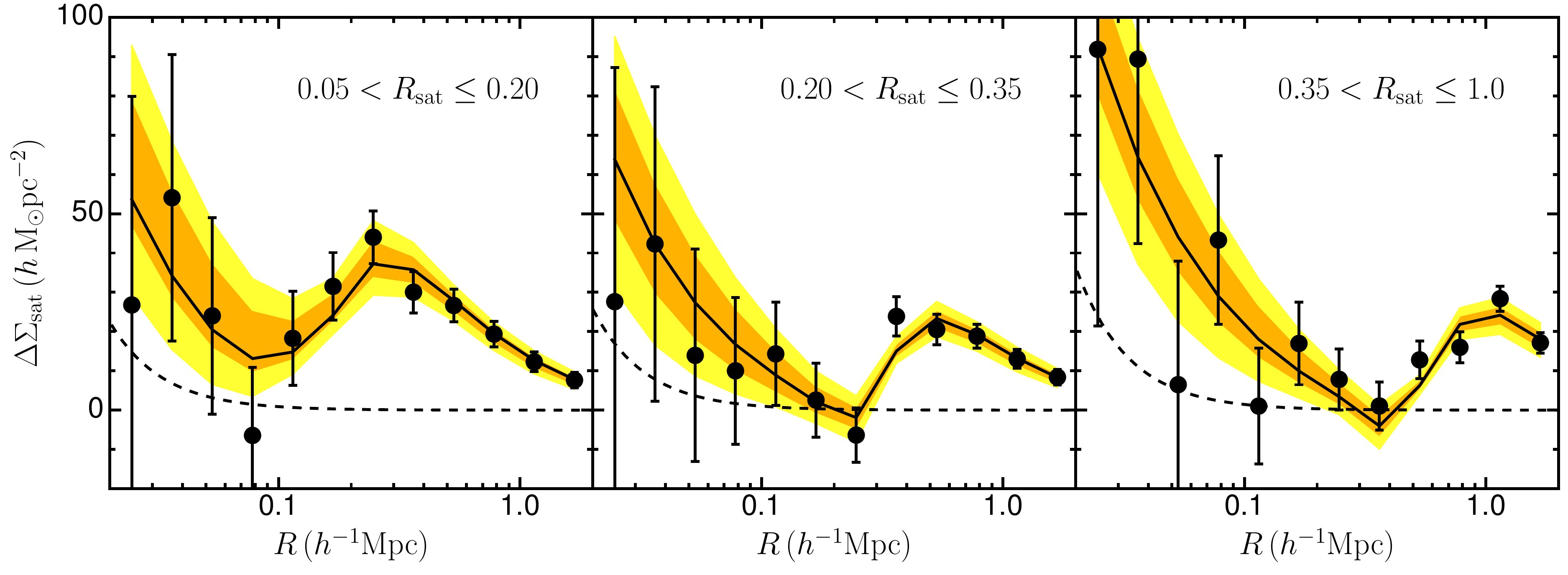}}
\caption{Excess surface density around satellite 
galaxies in the three radial bins summarized in \Cref{t:sample} and shown in the legends in units 
of $h^{-1}{\rm Mpc}$. Black points show lensing measurements around GAMA group satellites using 
KiDS data; errorbars correspond to the square root of the diagonal elements of the covariance 
matrix. The solid black line is the best-fit model where subhaloes are modelled as having NFW 
density profiles, and orange and yellow shaded regions mark 68\% and 95\% credible intervals, 
respectively. Dashed lines show the contribution of a point mass with a mass equal to the median 
stellar mass of each bin, which is included in the model.}
\label{f:esd_satellites}
\end{figure*}

\cite{pastormira11} studied the density profiles of subhaloes in the Millenium simulation 
\citep{springel05} and found that they are well fit by an NFW profile, with no evidence for 
truncation at any separation from the group centre. As discussed by \cite{hayashi03}, while tidal 
disruption removes mass preferentially from the outskirts, tidal heating causes the subhalo to 
expand after every orbit. The two effects compensate in terms of the density distribution such that 
a defined truncation radius cannot be discerned in subhaloes. We therefore model subhaloes as NFW 
profiles (\Cref{eq:NFW}). We assume the $c(M,z)$ relation of \cite{duffy08}; in analogy to 
\Cref{eq:cM}, we set $f_c^{\rm sub}=1$. To account for the baryonic contribution to the subhalo 
mass, we include a point mass in the centre with a mass equal to the median stellar mass for each 
bin (\Cref{t:sample}). Our model for the satellites therefore has a single free parameter per radial 
bin, namely $M_{\rm sub}(<r_{200})$.

For comparison, we also implement a theoretically-motivated model where subhaloes are tidally 
stripped by the host potential. In this model, a subhalo in a circular orbit is truncated at the 
radius at which the accelerations due to the tidal force from the host halo equals that arising 
from the gravitational force of the subhalo itself. This radius is given by
\begin{equation}\label{eq:rt}
 r_{\rm t} = \left[\frac{M_{\rm sub}(<r_{\rm t})}{\left(3-\partial\ln M/\partial\ln r\right)M_{\rm 
host}(<r_{\rm sat})}\right]^{1/3}r_{\rm sat}
\end{equation}
\citep{king62,binney87,mo10}, where, for an NFW profile,
\begin{equation}\label{eq:dlnMdlnr}
 \frac{\partial\ln M_{\rm NFW}}{\partial\ln r} = 
\frac{r^2}{\left(r_s+r\right)^2}\left[\ln\left(\frac{r_s+r}{r_s}\right) - 
\frac{r}{r_s+r}\right]^{-1}
\end{equation}
and $r_s$ is the scale radius of the host halo. Note that in \Cref{eq:rt} the truncation radius, 
$r_{\rm t}$, depends on the 3-dimensional distance to the group centre, $r_{\rm sat}$, which is not 
an observable. We draw 3-dimensional radii randomly from an NFW profile given the distribution of 
projected separations, $R_{\rm sat}$, for each bin. We additionally force $r_{\rm t} \leq r_{200}$, 
although the opposite rarely happens.

We model the truncation itself in a simple fashion, with an NFW profile instantaneously and 
completely stripped beyond $r_{\rm t}$,
\begin{equation}\label{eq:tNFW}
 \rho_t(r) = 
\begin{cases}
 \rho_{\rm NFW}(r) & r \leq r_{\rm t} \\
 0 & r > r_{\rm t}.
\end{cases}
\end{equation}
Note that, in addition to $r_{\rm t}$, this profile is defined \textit{mathematically} by the same 
parameters, $c$ and $M_{200}$, as a regular NFW, even though they are not well-defined 
physically; when referring to truncated models we report the proper physical masses, $M_{\rm 
sub} \equiv M_{\rm sub}(<r_{\rm t})$. We use the analytical expression for the ESD of the density 
profile given by \Cref{eq:tNFW} derived by \cite{baltz09}. In the leftmost panel of
\Cref{f:illustration_satellite} we show the ESD corresponding to such profile, compared with the 
ESD obtained assuming our fiducial NFW profile. The sharp truncation of the profile creates a glitch 
in the ESD around satellite galaxies at the radius of truncation which is continuous but 
non-differentiable and which, given our errorbars (cf.\ \Cref{f:esd_all}), has no impact on our 
results. The other panels show the effect of the three parameters describing the truncated subhalo 
density profile, \Cref{eq:tNFW} (the full NFW profile follows the same description but without the 
sharp cut at $r_{\rm t}$): as for the group profile, the mass and concentration affect the 
normalization and slope of the profile, respectively; the rightmost panel shows changes in $r_{\rm 
t}$ for the same subhalo mass \textit{within} $r_{\rm t}$, which is why the normalization of the 
different curves is different.

\section{Results}\label{s:results}

We show the ESD around satellites in each of the three radial bins in \Cref{f:esd_satellites}. 
Qualitatively, the signal looks similar to that of \Cref{f:esd_all}, and the features described in 
\Cref{s:lensing} are clearly seen in each of the panels. The dip in the signal close to the typical 
$\Rsat$ is smooth, as anticipated in \Cref{s:sigma_host}, and moves to higher $R$ with increasing 
$\Rsat$, as expected. As in \Cref{f:esd_all}, the errorbars correspond to the square root of 
the diagonal elements of the covariance matrix (see \Cref{ap:cov}).

After describing the fitting procedure in \Cref{s:fitting}, we summarize our constraints on group 
properties in \Cref{s:groupmasses} and on the satellite masses in \Cref{s:masses}. In 
\Cref{s:shmf} we carry out a proof-of-concept comparison of our results to predictions from 
semi-analytical models of subhalo statistics and we discuss the effect of contamination in the 
group sample in \Cref{s:contamination}.

\subsection{Fitting procedure}\label{s:fitting}

We fit the data in \Cref{f:esd_satellites} with the model described in \Cref{s:lensing} for each of 
the radial bins, using the median redshift of each galaxy sample, $\langle z_{\rm sat} \rangle$, 
as listed in \Cref{t:sample}. We use a single normalization $f_c^{\rm host}$ for the $c(M,z)$ 
relation of groups in the three bins. Our model therefore has seven free parameters: the three 
(weighted average) masses of the satellites, the three group masses, and a normalization to the 
$c(M,z)$ relation of \cite{duffy08} which applies to all groups across satellite radial bins.

We implement the model described above in a Markov Chain Monte Carlo (MCMC) using 
\texttt{emcee}\footnote{\url{http://dan.iel.fm/emcee/current/}} \citep{foreman13}, which is based 
on an affine-invariant ensemble sampler. This sampler works by using a number of `walkers' (in our 
case, a few hundred), each of which starts at a slightly different position in parameter space. 
Each step is drawn for each walker from a Metropolis-Hastings proposal based on the positions of 
all other walkers at the previous step \citep[see][for details about the algorithm]{goodman10}. The 
likelihood $\mathcal{L}$ is given by
\begin{equation}\label{eq:likelihood}
 \mathcal{L} = 
 \frac1{(2\pi)^{9/2}}\prod_{m=1}^3\prod_{n=1}^3\frac1{\sqrt{\lvert\mathbfss{C}_{mn}\rvert}}
 \exp\left[-\frac1{2}\boldsymbol{(O-E)}^T_m\mathbfss{C}^{-1}_{mn}\boldsymbol{(O-E)}_n\right],
\end{equation}
where $\boldsymbol{O}_m$ and $\boldsymbol{E}_m$ are the measurements and model predictions in 
radial bin $m$, respectively; $\mathbfss{C}^{-1}_{mn}$ is the element of the inverse covariance 
matrix that accounts for the correlation between radial bins $m$ and $n$; and 
$\lvert\mathbfss{C}_{mn}\rvert$ is the corresponding determinant. We therefore account for 
covariance both within and between radial bins in our MCMC. We assume flat, broad priors for all 
parameters, as listed in \Cref{t:results}.

The data are well fit by the model of \Cref{s:lensing}. The best-fit model is shown in 
\Cref{f:esd_satellites} and gives $\chi^2=24.7$ with 28 degrees of freedom, with a probability to 
exceed ${\rm PTE}=0.64$. Joint 2-dimensional posterior distributions for the seven free parameters 
are shown\footnote{We show and list the results in logarithmic space for convenience, but the 
analysis has been carried out in linear space and the reported uncertainties correspond to the 
uncertainties in linear space expressed on a logarithmic scale.}. Marginalized 
posterior estimates for all seven parameters, together with 68\% credible intervals, are reported 
in \Cref{t:results}, which also lists the stellar mass fractions, fractional satellite masses, and 
group mass-to-light ratios derived from the posterior mass estimates.

\subsection{Group masses and mass-concentration relation}\label{s:groupmasses}

Before discussing the results for the satellite galaxies, we explore the constraints on group 
masses and the group $c(M,z)$ relation. The masses of the same galaxy groups have been directly 
measured by \cite{viola15}, which provides a valuable sanity check of our estimates.

\begin{figure*}
 \centerline{\includegraphics[width=6in]{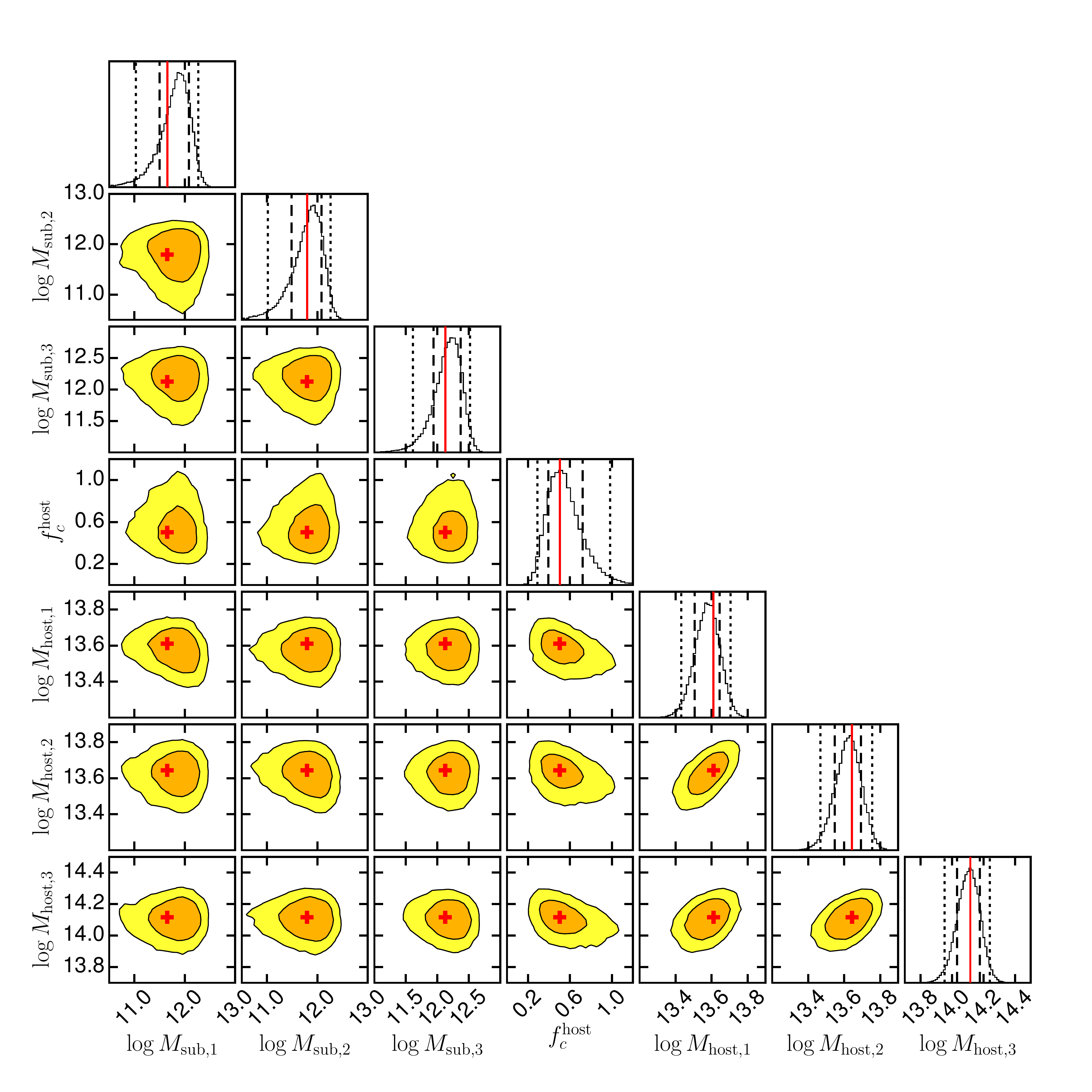}}
\caption{Joint 2-dimensional (lower off-diagonal panels, with contours at 68\% and 95\% joint 
credible regions) and marginalized 1-dimensional (diagonal panels) posterior distributions of free 
parameters of the model described in \Cref{s:lensing}, with subhaloes modelled with NFW density 
profiles. In the diagonal panels, black dashed and dotted lines mark marginalized 68\% and 95\% 
credible intervals, respectively, and vertical red solid lines mark the maximum likelihood estimate. 
Red crosses in off-diagonal panels show the joint best-fit values. All masses are in units of 
$h^{-1}{\rm M}_\odot$ and are numbered according to the radial bin to which they correspond.}
\label{f:corner}
\end{figure*}

We find that the normalization of the $c(M,z)$ relation is significantly lower than the fiducial 
\cite{duffy08} relation, $f_c^{\rm host}=0.53_{-0.14}^{+0.19}$ (where the fiducial value is 
$f_c^{\rm host}=1$). This normalization implies concentrations $c\approx3$ for these groups. For 
comparison, using the same parameterization as we do, \cite{viola15} measured $f_c^{\rm 
host}=0.84_{-0.23}^{+0.42}$. Our smaller errorbars are due to the fact that we do not account for 
several nuisance parameters considered by \cite{viola15} in their halo model implementation. Most 
notably, accounting for miscentring significantly increases the uncertainty on the concentration, 
since both affect $\Delta\Sigma$ at similar scales \citep{viola15}. Indeed, when they do not account 
for miscentring, \cite{viola15} measure $f_c^{\rm host}=0.59_{-0.11}^{+0.13}$, consistent with our 
measurement both in the central value and the size of the errorbars. While this means that our 
estimate of $f_c^{\rm host}$ is biased, accounting for extra nuisance parameters such as miscentring 
is beyond the scope of this work; our aim is to constrain satellite masses and not galaxy group 
properties. As shown in \Cref{f:corner}, $f_c^{\rm host}$ is not correlated with any of the other 
model parameters and therefore this bias in $f_c^{\rm host}$ does not affect our estimates of the 
satellite masses. 

Group masses are consistent with the results from \cite{viola15} (with the same caveat that 
the small errorbars are an artifact produced by our simplistic modelling of the host groups).
Specifically, our average mass-to-light ratios follow the mass-luminosity relation found by 
\cite{viola15}, $M_{\rm 200} \propto L_{200}^{1.16\pm0.13}$. As shown in \Cref{f:corner}, group 
masses are slightly correlated because they are forced to follow the same mass-concentration 
relation determined by \Cref{eq:cM}. Groups in the third bin are on average $\sim3.4\pm0.8$ times 
more massive than groups in the first radial bin. This is a selection effect, arising because groups 
in each bin must be big enough to host a significant number of satellites at the characteristic 
radius of each bin. For example, groups in the first radial bin have\footnote{Throughout, we quote 
masses and radii for a given radial bin by adding an index from 1 to 3 to the subscript of each 
value.} in \Cref{f:corner} $\log \langle M_{\rm 
host,1}/(h^{-1}{\rm M}_\odot) \rangle=13.46_{-0.06}^{+0.06}$ and $\langle c_1 \rangle\approx3.3$, 
which implies a scale radius $\langle r_{s,1}\rangle =0.19\,h^{-1}{\rm Mpc}$, beyond which the 
density drops as $\rho \propto r^{-3}$ (cf.\ \Cref{eq:NFW}). The average 3-dimensional distance of 
satellites to the group centre (see \Cref{s:sigma_sat}) in the third radial bin is $\langle r_{\rm 
sat,3} \rangle=0.46\,h^{-1}{\rm Mpc}$. At this radius, the average density in groups in the first 
radial bin is seven times smaller than at $\langle r_{s,1} \rangle$.

As mentioned above, our simplistic modelling of groups does not affect the posterior satellite 
masses significantly. Therefore it is sufficient that our group masses are consistent with the 
results of \cite{viola15}, and we do not explore more complex models for the group signal. For a 
more thorough modelling of the lensing signal of groups in the KiDS-GAMA overlap region, see 
\cite{viola15}.

\subsection{The masses of satellite galaxies}\label{s:masses}

\begin{figure}
 \centerline{\includegraphics[width=3.4in]{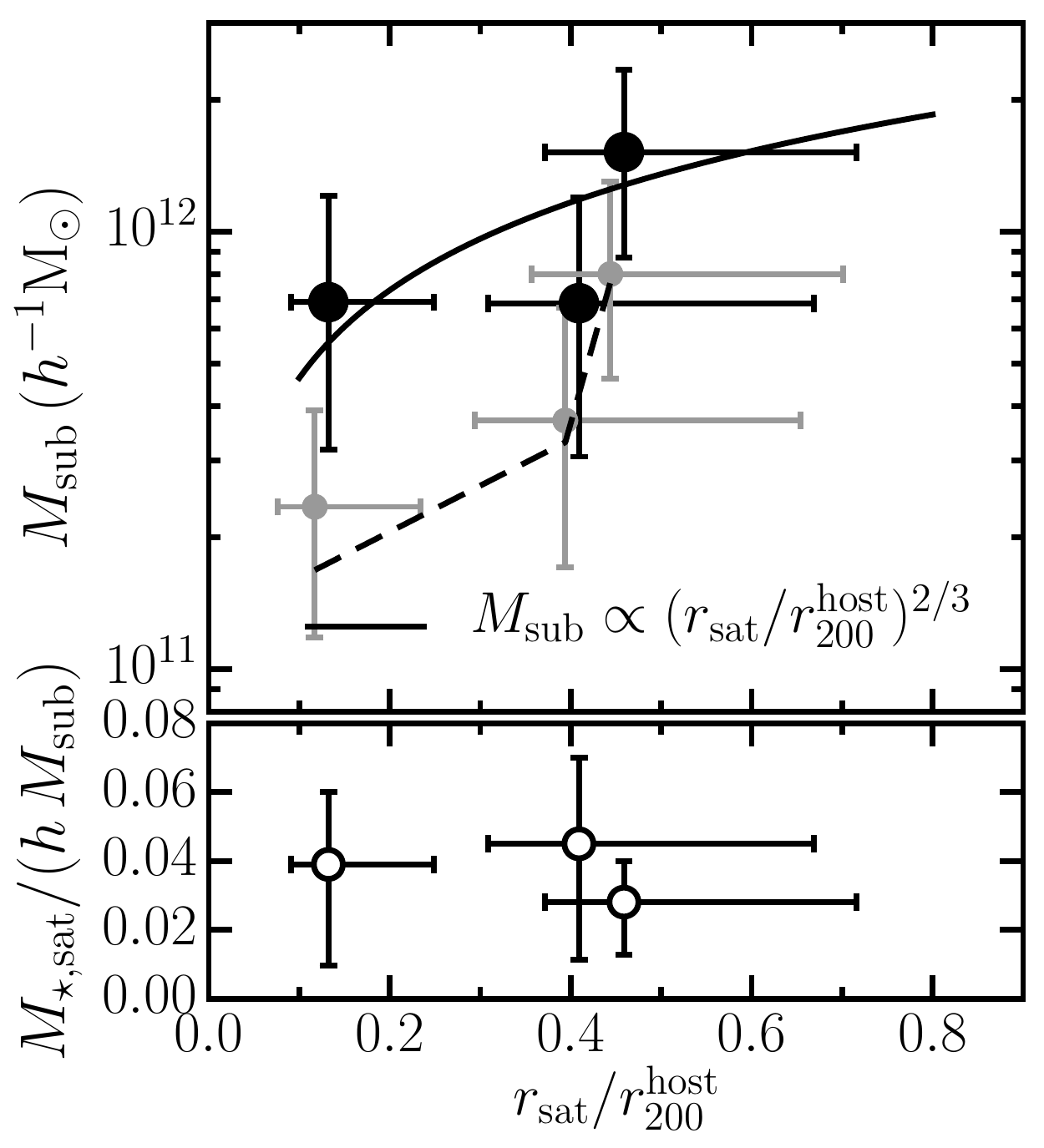}}
\caption{\textit{Top:} Marginalized posterior mass estimates of satellite galaxies from the full NFW 
(black, large points) and truncated NFW (grey, small points) models, and the dashed black line shows 
the NFW masses within the same truncation radii for comparison. Horizontal errorbars are 68\% ranges 
in (3-dimensional) $r_{\rm sat}/r_{200}$ per bin. The black solid line shows the radial dependence 
of subhalo mass predicted by the numerical simulations of \citet{gao04} with an arbitrary 
normalization.
\textit{Bottom:} Stellar-to-total mass ratios in each bin.}
\label{f:msat_r}
\end{figure}

\begin{table*}
\begin{center}
\caption{Priors, marginalized posterior estimates and derived parameters of the satellites and host 
groups in the three radial bins. All priors are uniform in linear space in the quoted range.  We use 
medians as central values and all uncertainties are 68\% credible intervals. The normalization of 
the group $c(M,z)$ relation, $f_c^{\rm host}$, is the same for the three radial bins. The best-fit 
model has $\chi^2=24.7$ with 28 degrees of freedom (PTE = 0.64).}
\label{t:results}
\begin{tabular}{c|c c c c c}
\hline\hline
Parameter & Units & Prior & Bin 1 & Bin 2 & Bin 3 \\
\hline
$\log M_{\rm sub}$ & $h^{-1}{\rm M}_\odot$ & $[7,13]$ & $11.84_{-0.34}^{+0.24}$ & 
$11.84_{-0.35}^{+0.24}$ & $12.18_{-0.24}^{+0.19}$ \\[0.3ex]
$f_c^{\rm host}$ & 1 & $[0,2]$ & $0.53_{-0.14}^{+0.19}$ & \checkmark & \checkmark \\[0.3ex]
$\log M_{\rm host}$ & $h^{-1}{\rm M}_\odot$ & $[10,15]$ & $13.58_{-0.07}^{+0.07}$ & 
$13.62_{-0.08}^{+0.07}$ & $14.11_{-0.07}^{+0.07}$ \\[0.3ex]
\hline
\multicolumn{6}{c}{Derived Parameters} \\
$\langle M_{\rm \star,sat}\rangle / \langle M_{\rm sub} \rangle$ & $h^{-1}$ & -- &
 $0.04_{-0.03}^{+0.02}$ & $0.04_{-0.03}^{+0.02}$ & $0.03_{-0.02}^{+0.01}$ \\[0.3ex]
$\langle M_{\rm sub}/M_{\rm host} \rangle$ & 1 & -- &
 $0.018_{-0.010}^{+0.014}$ & $0.016_{-0.009}^{+0.013}$ & $0.012_{-0.005}^{+0.007}$ \\[0.3ex]
$\langle M_{\rm host}\rangle / \langle L_{\rm host} \rangle$ & $h\,{\rm M}_\odot/{\rm L}_\odot$ & 
-- & $300_{-45}^{+49}$ & $265_{-42}^{+46}$ & $386_{-61}^{+66}$ \\[0.3ex]
\hline
\end{tabular}
\end{center}
\end{table*}

We detect the signal from satellites with significances $>$99\% in all three radial bins. 
Satellite masses are consistent across radial bins. We show the marginalized posterior estimates 
and 68\% credible intervals in \Cref{f:msat_r} as a function of 3-dimensional group-centric 
distance, $\rsat$ (in units of the group radius $r_{200}$).

\Cref{f:msat_r} also shows the subhalo mass as a function of 3-dimensional separation from the group 
centre found in numerical simulations by \cite{gao04}. Note that we compare here only the 
trend with radius, not the normalization. Fitting a power law, $M_{\rm sub} \propto (r_{\rm 
sat}/r_{200})^a$, to the data in \Cref{f:msat_r} we find $a=0.3\pm0.5$ (ignoring horizontal 
errorbars), consistent with the trend predicted by \cite{gao04} but also with no dependence on 
group-centric distance. The bottom panel shows the average stellar mass fractions, which are also 
consistent with each other, $\langle M_{\rm \star,sat}/M_{\rm sub}\rangle \sim0.04\,h^{-1}$.

We also show in \Cref{f:msat_r} the results obtained for the truncated theoretical model. The 
difference between each pair of points depends on the posterior $r_{\rm t}$ estimated in each 
bin through \Cref{eq:rt}. Specifically, we estimate $\langle r_{\rm 
t}\rangle=\{0.04_{-0.01}^{+0.02},0.06_{-0.02}^{+0.03},0.09_{-0.02}^{+0.04}\}\,h^{-1}$Mpc. We 
remind the reader that these are theoretical predictions from \Cref{eq:rt} rather than 
observational results. For comparison, we also show in \Cref{f:msat_r} the masses obtained by 
integrating the posterior NFW models up to said truncation radii, shown by the dashed line. 
These masses are fully consistent with the truncated model, implying that the difference 
between the black and grey points (which show $M_{\rm sub}(<\!r_{200})$ and $M_{\rm sub}(<\!r_{\rm 
t})$, respectively) in \Cref{f:msat_r} is only a matter of presentation; the data cannot 
distinguish between these two models.

After we submitted this work, \cite{li15} presented similar, independent satellite lensing 
measurements. They used $\sim$7,000 satellites in the redMaPPer galaxy group catalogue 
\citep{rykoff14} with background sources from CS82 and also measured the lensing signal in three 
bins in projected radius. They find comparable constraints that are consistent with ours.

\subsection{The average subhalo mass}\label{s:shmf}

We can link the results presented in \Cref{s:masses} to predictions from numerical simulations. 
Comparisons of the satellite populations of observed galaxies (or groups) provide valuable insights 
as to the relevant physical processes that dominate galaxy formation, as highlighted by the well 
known `missing satellites' \citep{klypin99,moore99} and `too big to fail' \citep{boylan11} problems, 
which suggest either that our Universe is not well described by a $\Lambda$CDM cosmology, or that 
using numerical simulations to predict observations is more complicated than anticipated. While the 
former may in fact be true, the latter is now well established, as the formation of galaxies inside 
dark matter haloes depends strongly on baryonic physics not included in $N$-body simulations, and 
the influence of baryons tends to alleviate these problems \citep{zolotov12}.

Here we specifically compare the average subhalo-to-host mass ratio, $\psi \equiv M_{\rm 
sub}/M_{\rm host}$, to $\Lambda$CDM predictions through the subhalo mass function, which describes 
the mass distributions of subhaloes for a given dark matter halo mass. In numerical simulations, the 
resulting subhalo mass function is a function only of $\psi$ 
\citep[e.g.,][]{vandenbosch05,jiang14}. 
As summarized in \Cref{t:results}, we find typical subhalo-to-host mass ratios in the range 
$\langle\psi\rangle\sim0.015$, statistically consistent across group-centric distance. We obtain 
these values by taking the ratio $M_{\rm sub}/M_{\rm host}$ at every evaluation in the MCMC. For 
comparison, the values we obtain using the truncated model are $\langle\psi_{\rm 
tNFW}\rangle\approx0.005$, also consistent across radial bins.

We compare our results to the analytical \emph{evolved} (that is, measured after the subhaloes have 
become satellites of the host halo, as opposed to one measured at the time of infall) subhalo mass 
function proposed by \cite{vandenbosch05},
\begin{equation}\label{eq:shmf}
 \frac{{\rm d}N}{{\rm d}\psi} \propto \frac1{\psi}\left(\beta/\psi\right)^\alpha 
\exp\left(-\psi/\beta\right),
\end{equation}
where $\alpha=0.9$ and $\beta=0.13$, and calculate the average subhalo-to-host mass ratio,
\begin{equation}
 \langle\psi\rangle =
  \left[\int_{\psi_{\rm min}}^{\psi_{\rm max}}\frac{{\rm d}N}{{\rm d}\psi}{\rm d}\psi\right]^{-1} 
   \int_{\psi_{\rm min}}^{\psi_{\rm max}} \psi\frac{{\rm d}N}{{\rm d}\psi}{\rm d}\psi,
\end{equation}
where $\psi_{\rm min}\approx10^{-3}$ is approximately the minimum fractional satellite mass we 
observe given the results of \Cref{s:masses}, and $\psi_{\rm max}=1$ is the maximum fractional 
satellite mass by definition. Integrating in this range gives $\langle\psi\rangle=0.0052$.

There are many uncertainties involved in choosing a $\psi_{\rm min}$ representative of our sample, 
such as survey incompleteness and the conversion between stellar and total mass; we defer a proper 
modelling of these uncertainties to future work. For reference, changing $\psi_{\rm min}$ by a 
factor 5 modifies the predicted $\langle\psi\rangle$ by a factor $\sim$3. Considering the 
uncertainties involved, all we can say at present is that our results are consistent with 
$\Lambda$CDM predictions.

\subsection{Sensitivity to contamination in the group catalogue}
\label{s:contamination}

Two sources of contamination in the group catalogue have been neglected in this analysis. The 
spectroscopic group satellite catalogue used in this work has a high, but not 100\%, purity. For 
groups with $N_{\rm FoF}\geq5$ the purity approaches 90\%; groups with fewer members 
have significantly lower purity \citep{robotham11}. \cite{li13} have shown that a contamination 
fraction of 10\% in the satellite sample would lead to a $+15\%$ bias in the inferred satellite 
masses, well within the reported uncertainties (which amount to up to a factor two).

The second source of contamination is the misidentification of the central galaxy in a group, 
such that the true central galaxy would be included in our satellite sample. This effect is similar 
to that explained above, except that contaminating galaxies now reside in particularly massive 
halos (namely, the groups themselves). Based on comparisons to GAMA mock galaxy catalogues, 
\cite{robotham11} found that the fraction of BCGs correctly identified with the central galaxy of 
dark matter halos is around $70-75\%$ for groups with $N_{\rm FoF}\geq5$. \cite{viola15} have 
directly measured the offset probability of BCGs from the true minimum of the potential well. They 
found that the BCG is as good a proxy for the centre as the iterative centre of \cite{robotham11}, 
which according to mock group catalogues are well centred in $\sim90\%$ of the groups. There 
are very few groups with $N_{\rm FoF}\gg5$ \citep{robotham11}, and therefore the lensing signal of a 
central galaxy in our sample would probably have $\Delta\Sigma(R\approx0.05\,h^{-1}{\rm 
Mpc})\approx100\,h\,\mathrm{M_\odot\,pc^{-2}}$ \citep[see Figure 7 of][]{viola15}. If we assume 
(conservatively) that 20\% of the BCGs do not correspond to the central galaxy in their groups, 
then $0.20/7=3\%$ (where $\langle N_{\rm FoF}\rangle\sim7$, cf.\ \Cref{t:sample}) of our satellites 
would be central galaxies. Therefore the total signal in the inner regions 
($R\approx0.05\,h^{-1}{\rm Mpc}$) would be $\Delta\Sigma_{\rm tot}=0.03\times100 + 
0.97\times\Delta\Sigma_{\rm sub}^{\rm true}\simeq40\,h\,\mathrm{M_\odot\,pc^{-2}}$, which yields 
$\Delta\Sigma_{\rm sub}^{\rm true}=38\,h\,\mathrm{M_\odot\,pc^{-2}}$. Therefore central galaxy 
misidentification induces a $+5\%$ bias on the signal, which implies roughly a $+15\%$ bias on the 
mass.

Together, these two effects add up to a $\sim\!20-25\%$ bias in our satellite mass estimates. 
Such a bias is safely within our statistical uncertainties. Therefore our results are insensitive to 
plausible levels of contamination in the group catalogue, both from satellites that are not really 
group members and from misidentified central galaxies.

\section{Conclusions}\label{s:conclusions}

We used the first 100 \sqdeg\ of optical imaging from KiDS to measure the excess surface mass 
density around spectroscopically confirmed satellite galaxies from the GAMA galaxy group catalogue. 
We model the signal assuming NFW profiles for both host groups and satellite galaxies, including the 
contribution from the stellar mass for the latter in the form of a point source. Taking advantage of 
the combination of statistical power and high image quality, we split the satellite population into 
three bins in projected separation from the group centre, which serves as a (high-scatter) proxy for 
the time since infall. We fit the data with a model that includes the satellite and group 
contributions using an MCMC (see \Cref{s:lensing} and \Cref{f:esd_satellites}), fully accounting for 
the data covariance. As a consistency check, we find group masses in good agreement with the weak 
lensing study of GAMA galaxy groups by \cite{viola15}, even though we do not account for effects 
such as miscentring or the contribution from stars in the BCG.

This model fits the data well, with $\chi^2/{\rm d.o.f.}=0.88$ (${\rm PTE}=0.64$). We are able to 
constrain total satellite masses to within $\sim0.3$ dex or better. Given these uncertainties, 
the estimated masses are insensitive to the levels of contamination expected in the group 
catalogue. Satellite galaxies have similar masses across group-centric distance, consistent with 
what is found in numerical simulations (accounting for the measured uncertainties). Satellite masses 
as a function of group-centric distance are influenced by a number of effects. Tidal stripping acts 
more efficiently closer to the group centre, while dynamical friction makes massive galaxies sink to 
the centre more efficiently, an effect referred to as mass segregation \citep[e.g.,][]{frenk96}. In 
addition, by binning the sample in (projected) group-centric distance we are introducing a selection 
effect such that outer bins include generally more massive groups, which will then host more massive 
satellites on average. Future studies with increased precision may be able to shed light on the 
interplay between these effects by, for instance, selecting samples residing in the same host groups 
or in bins of stellar mass. 

As a proof of concept, we compare our results to predictions from $N$-body simulations. These 
predict that the subhalo mass function is a function only of the fractional subhalo mass, $\psi 
\equiv M_{\rm sub}/M_{\rm host}$. Our binning in satellite group-centric distance produces a 
selection effect on host groups, such that each bin probes a (slightly) different group population, 
which allows us to test such prediction. The average fractional mass in all three bins is consistent 
with a single value (within large errorbars), $\langle\psi\rangle\sim0.015$. This is broadly 
consistent with the predictions of numerical simulations. We anticipate that weak lensing of 
satellite galaxies will become an important tool to constrain the physical processes incorporated in 
semi-analytic models of galaxy formation and, ultimately, hydrodynamical simulations.

\section*{acknowledgments}

We thank Martin Eriksen, Joachim Harnois-D\'eraps and Thomas Kitching for useful comments on the 
manuscript. We are grateful to Matthias Bartelmann for being our external blinder, revealing which 
of the four catalogues analysed was the true unblinded catalogue at the end of this study.

C.S., H.Ho., M.V., A.C.\ and C.H.\ acknowledge support from the European Research Council under FP7 
grant numbers 279396 (CS, HH and MV) and 240185 (AC, CH).
H.Ho., M.C., M.V.\ and J.d.J.\ acknowledge support from the Netherlands Organisation for Scientific
Research (NWO) grant numbers 639.042.814 (HH and MC) and 614.001.103 (MV) and 614.061.610 (JdJ).
E.v.U.\ acknowledges support from a grant from the German Space Agency DLR and from an STFC Ernest 
Rutherford Research Grant, grant reference ST/L00285X/1.
H.Hi.\ is supported by the DFG Emmy Noether grant Hi 1495/2-1.
B.J.\ acknowledges support by an STFC Ernest Rutherford Fellowship, grant reference ST/J004421/1.
R.N.\ acknowledges support from the German Federal Ministry for Economic Affairs and Energy (BMWi) 
provided via DLR under project No.\ 50QE1103.
P.S.\ is supported by the Deutsche Forschungsgemeinschaft in the framework of the TR33 `The Dark 
Universe'.
G.V.K.\ acknowledges financial support from the Netherlands Research School for Astronomy (NOVA) 
and Target. Target is supported by Samenwerkingsverband Noord Nederland, European fund for regional 
development, Dutch Ministry of economic affairs, Pieken in de Delta, Provinces of Groningen and 
Drenthe.

Based on data products from observations made with ESO Telescopes at the La Silla Paranal 
Observatory under programme IDs 177.A-3016, 177.A-3017 and 177.A-3018, and on data products 
produced by Target/OmegaCEN, INAF-OACN, INAF-OAPD and the KiDS production team, on behalf of the 
KiDS consortium. OmegaCEN and the KiDS production team acknowledge support by NOVA and NWO-M 
grants. Members of INAF-OAPD and INAF-OACN also acknowledge the support from the Department of 
Physics \& Astronomy of the University of Padova, and of the Department of Physics of Univ. 
Federico II (Naples).

GAMA is a joint European-Australasian project based around a spectroscopic campaign using the 
Anglo-Australian Telescope. The GAMA input catalogue is based on data taken from the Sloan Digital 
Sky Survey and the UKIRT Infrared Deep Sky Survey. Complementary imaging of the GAMA regions is 
being obtained by a number of independent survey programs including GALEX MIS, VST KiDS, VISTA 
VIKING, WISE, Herschel-ATLAS, GMRT and ASKAP providing UV to radio coverage. GAMA is funded by the 
STFC (UK), the ARC (Australia), the AAO, and the participating institutions. The GAMA website is 
\url{http://www.gama-survey.org/}.

This work has made use of the \textsc{python} packages \textsc{numpy}, \textsc{scipy} and 
\textsc{IPython} \citep{perez07}. Plots have been produced with \textsc{matplotlib} 
\citep{hunter07}.

\textit{Author contributions:}  All authors contributed to the development and writing of this 
paper. The authorship list reflects the lead authors (CS, MC, HH) followed by two alphabetical 
groups. The first alphabetical group includes those who are key contributors to both the scientific 
analysis and the data products. The second group covers those who have made a significant 
contribution either to the data products or to the scientific analysis.

\begin{appendix}
 
\section{Full satellite lensing correlation matrix and the contribution from sample 
variance}\label{ap:cov}

\begin{figure*}
 \centerline{\includegraphics[height=4in]{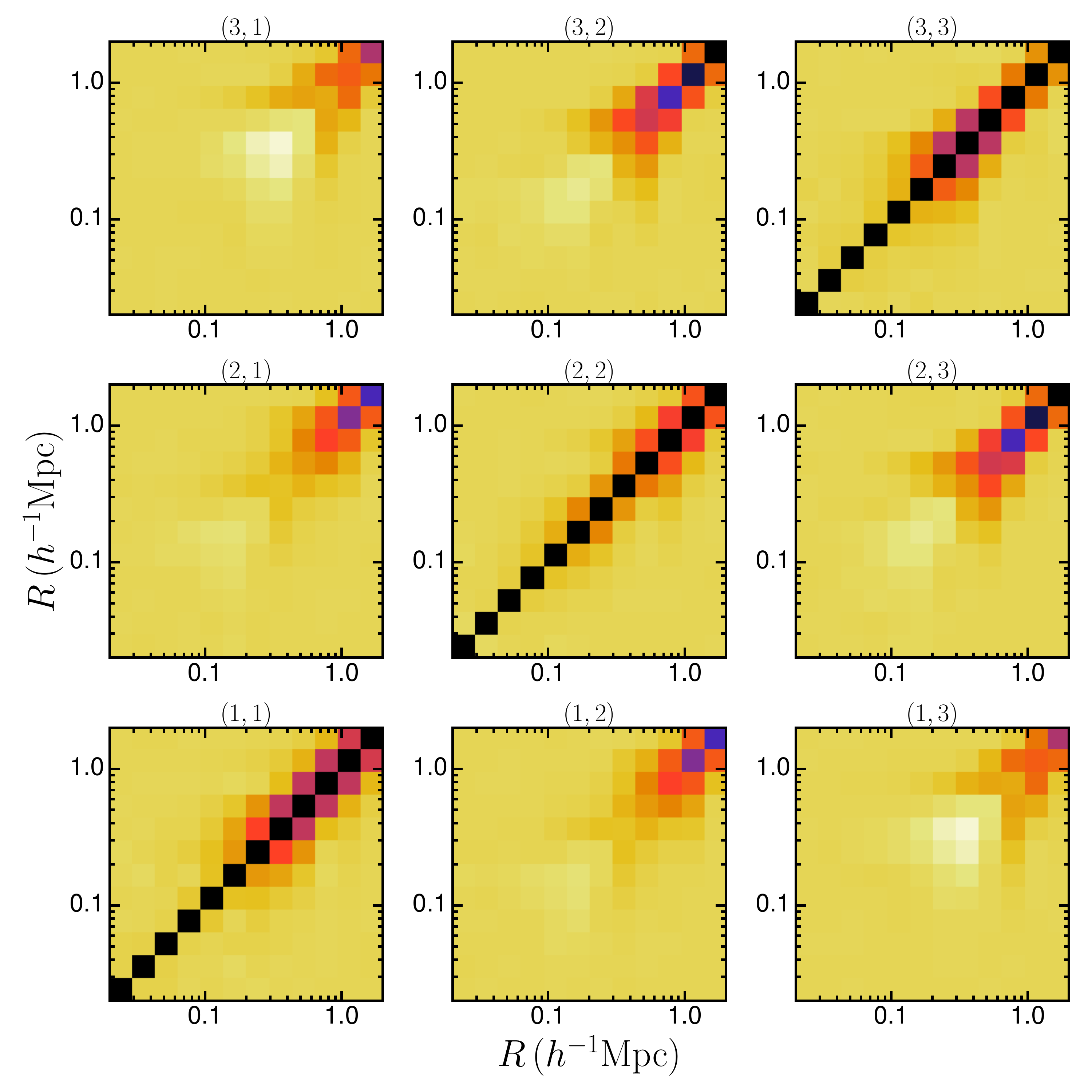}
             \includegraphics[height=4in]{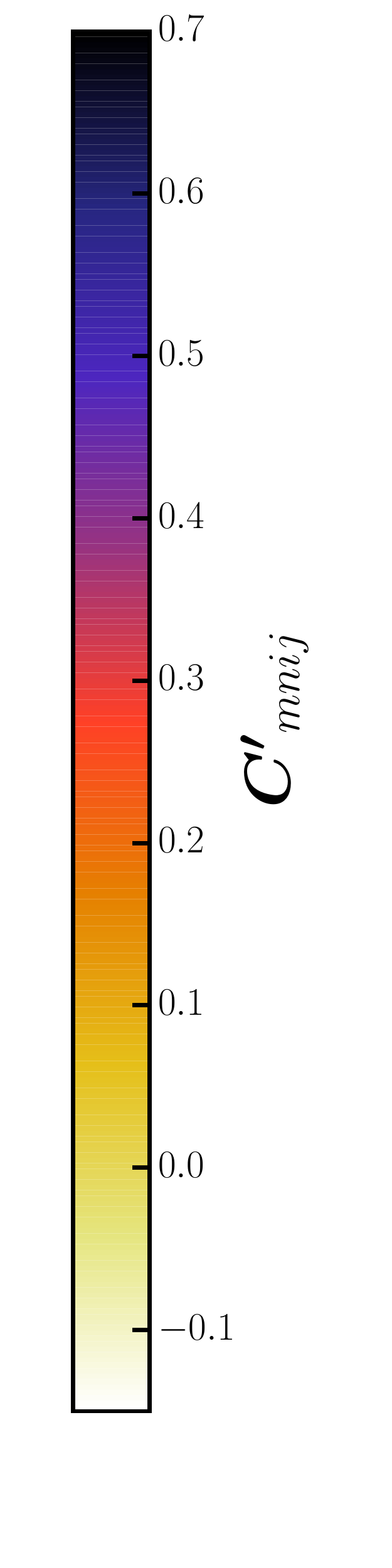}}
\caption{Full satellite lensing correlation matrix within and between radial bins as 
shown by the label at the top of each plot.}
\label{f:cov}
\end{figure*}

As mentioned in \Cref{s:signal}, we calculate the covariance matrix directly from the data 
including only the contribution from shape noise \citep[see Section 3.4 of][]{viola15}. In 
\Cref{f:cov} we show the corresponding correlation matrix, defined as
\begin{equation}\label{eq:corr}
 \boldsymbol{C'}_{mnij}=\frac{\boldsymbol{C}_{mnij}} 
{\sqrt{\boldsymbol{C}_{mmii}\boldsymbol{C}_{nnjj}}},
\end{equation}
where $\boldsymbol{C}_{mnij}$ is the covariance between the $i$-th and $j$-th elements of radial 
bins $m$ and $n$, respectively (where $m,n=1,2,3$). In reality the lensing covariance also includes 
a contribution from sample (`cosmic') variance, but we have ignored it in our analysis. Below we 
justify this decision.

The contribution from sample variance can in principle be estimated by bootstrapping the lensing 
signal over individual KiDS fields. However, there are two caveats to this approach. First, the 101 
KiDS fields used here do not produce enough independent bootstrap samples to properly estimate the 
full covariance matrix for our satellite samples, which is a symmetric $36\times36$ matrix 
(containing 648 independent elements) including sample variance for the three radial bins. 
Second, using single KiDS fields as bootstrap elements means that the elements are not 
truly independent from each other, because lenses in one field do contribute to signal in 
neighbouring fields. In fact, we calculate the lensing signal of each galaxy including background 
galaxies in neighbouring fields.

The latter point is not crucial for our analysis since, as shown in \Cref{f:esd_satellites}, the 
signal produced by satellite subhaloes is confined to the smallest scales, 
$R\lesssim0.3\,h^{-1}{\rm 
Mpc}$. Therefore, we can estimate the relative contribution from sample variance to the covariance 
matrix by comparing the diagonal sub-panels of the covariance matrices estimated directly from the 
data (the `analytical' covariance) and by bootstrapping over KiDS fields. Note that the bootstrap 
covariance also accounts for shape noise in addition to sample variance. Therefore the ratio 
between the bootstrap and analytical covariances is a measure of the relative contribution of 
sample variance to the satellite lensing covariance. It should be noted, however, that the 
bootstrap covariance can be biased high by as much as 40\% \citep{norberg09}.

\begin{figure}
 \centerline{\includegraphics[width=3in]{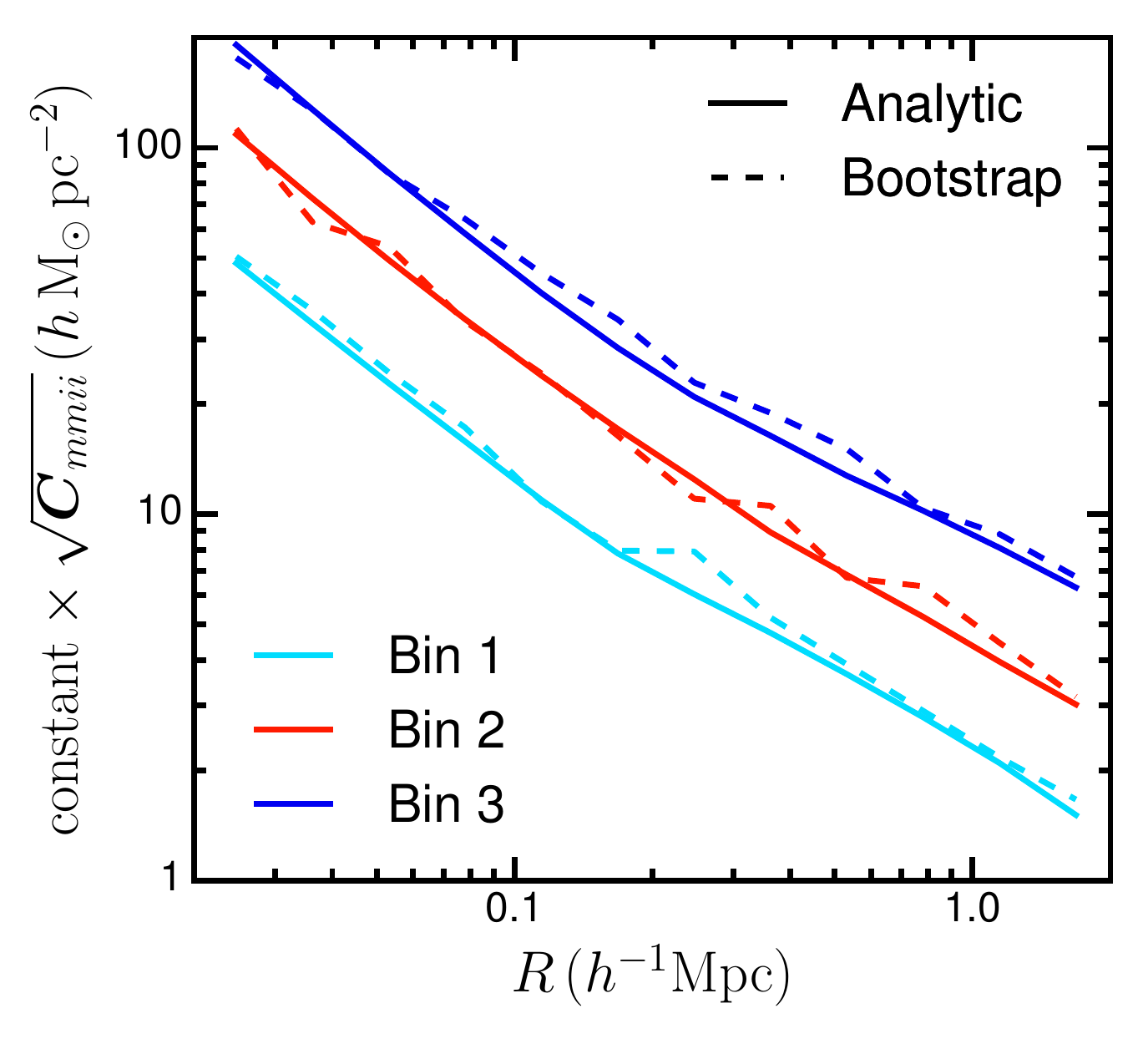}}
\caption{Comparison of the variances calculated analytically (solid), which account only for shape 
noise, and by bootstrapping (dashed), which also account for sample variance, for the diagonal 
sub-panels of the covariance matrix as per \Cref{f:cov}. The red and blue lines have been 
offset vertically for clarity.}
\label{f:compare_variances}
\end{figure}

We show this comparison in \Cref{f:compare_variances} for each of the three radial bins, where we 
compare $\sqrt{\boldsymbol{C}_{mmii}}$ estimated from the analytic (i.e., data) and bootstrap 
covariances. Both methods lead to similar values up to the largest angular separations. There is 
a hint of a nonzero contribution from sample variance at scales $R>0.3\,h^{-1}{\rm Mpc}$, where the 
bootstrap variance is $\sim10\%$ larger than the analytical variance. As stated above, the 
satellite 
contribution to $\Delta\Sigma$ is confined to scales smaller than these. We conclude that, for the 
purpose of this work, we can safely ignore sample variance.

\end{appendix}

\end{document}